\documentclass[%
 reprint,
nofootinbib,
 amsmath,amssymb,
 aps,
]{revtex4-2}

\usepackage{graphicx}
\usepackage{dcolumn}
\usepackage{bm}

\usepackage{tikz}
\usetikzlibrary{snakes}
\usetikzlibrary{calc,decorations.markings}
\usepackage[compat=1.1.0]{tikz-feynman}
\usepackage{subcaption}
\usepackage{feynmp-auto}
\usepackage{hyperref}
\usepackage{cleveref}
\usepackage{float}
\usepackage{physics}
\usepackage{pgfplots}
\usepackage{lipsum}
\usepackage[colorinlistoftodos]{todonotes}
\Crefname{equation}{Eq.}{Eqs.}
\Crefname{figure}{Fig.}{Figs.}
\Crefname{definition}{Def.}{Defs.}

\newcommand{\BigO}[1]{\mathcal{O}( #1 )}

\newenvironment{rcases}
  {\left.\begin{aligned}}
  {\end{aligned}\right\rbrace}

\begin{document}

\preprint{APS/123-QED}

\title{The Quantum Vacuum of Spacetime with a Fundamental Length}

\author{Paul C.W. Davies}
\email{paul.davies@asu.edu}
\author{Philip Tee}
\altaffiliation[Also at ]{p.tee@sussex.ac.uk, \\Department of Informatics, \\University of Sussex, Falmer, UK.}
\email{ptee2@asu.edu}
\affiliation{The Beyond Center for Fundamental Science,  Arizona State University, Tempe AZ 85287, USA}

\date{\today}

\begin{abstract}
A quantum theory of gravity implies a fine-grained structure of spacetime, which can be conveniently modeled as some form of ‘pixelation’ at the Planck scale, with potentially observable consequences. 
In this work, we build upon previous results to investigate the effect of pixelation on the quantum vacuum, making use of the framework of Doubly Special Relativity (DSR). 
At the center of the DSR approach is an observer dependent length scale, defining the pixelation of spacetime.
A key feature of quantum field theory in DSR is the dispersive nature of the vacuum state and the associated appearance of curvature in momentum space.
As a result, the standard treatment of the renormalized stress-energy-momentum tensor acquires correction terms. 
As an illustration, we present here a calculation of the thermal vacuum and modified Casimir effect, using both modified propagators and momentum measures. 
We choose a consistent choice of momentum space metric that both generates the modified dispersion relations we use and preserves the Lorentz invariant character of the results obtained.
Put together this constitutes a consistent calculation framework we can apply to other more complex scenarios.
\end{abstract}

\maketitle


\section{Introduction and Background}
\label{sec:introduction}

A familiar assumption is that spacetime is continuous, and the majority of contemporary theories of physics are described 
 using mathematics that assumes continuity.
However, even in classical physics the idea of a point particle such as an electron inevitably leads to divergences, and in quantum fields these problems become more serious.
With electrodynamics (and the nuclear forces), renormalization is successful at taming such divergences in a systematic fashion, but attempts to directly quantize gravity lead to a catastrophically divergent theory.
Could it be that the fundamental assumption of spacetime continuity needs to be reevaluated?

It is often speculated that the existence of a quantum theory of gravity (QG) will naturally lead to the idea that geometry itself is quantized \cite{hossenfelder2013minimal,wheeler2018information} and therefore ceases to be continuous.
There have been attempts to formulate theories of QG that do not require spacetime continuity such as Loop Quantum Gravity \cite{rovelli2014covariant}, Combinatorial Gravity \cite{trugenberger2015quantum,tee2020dynamics}, Causal Dynamical Triangulation \cite{dowker2006causal} and Quantum Geometry \cite{ambj1997quantum}, but none have been universally accepted or experimentally verified.
If spacetime is discrete, there would have to exist an observer-independent fundamental length scale usually expected to be the Planck length $l_p=1.616 \times 10^{-35}\text{m}$, although the precise value could be different.
In the absence of a full theory of quantum gravity (QG), is it possible to investigate leading order effects such a fundamental length scale would have on a quantized field?
If we can arrive at a result which is testable, or at least  one that surfaces logical consistency issues, this may throw light on the existence of a fine-grain structure or pixelation to spacetime.
The importance of a null result in such an investigation would be to cast doubt on geometry being quantized at the highest energy scales, and by implication perhaps bring into question whether gravity is in fact a quantum theory.
Given that experimental tests of QG are currently out of reach \cite{amelino2005introduction}, this provides an alternative route to establishing the existence of a theory of QG.
This is perhaps one of the most important open problems in theoretical physics.

Doubly Special Relativity (DSR), provides such a framework in which to study the effects of a fundamental length.
The key merit of DSR is that it is able to incorporate a fundamental length that is observer independent \cite{amelino2001testable}.
The approach rests upon modifications to the underlying group structure of the special theory of relativity \cite{amelino2002relativity,amelino2002coproduct}, by deforming the generators of the Poincar{\'e} group.
This has the effect of modifying energy-momentum dispersion relations, but crucially also the underlying geometry of momentum space $\mathcal{P}$.
Such modifications are not unique to DSR, for example other approaches to a quantum theory of gravity (QG), such as Ho{\u r}ava-Lifshitz \cite{hovrava2009quantum} also propose that dispersion relations are modified.
We do not intend to cover the details of DSR here, but the key point is that we can take an approximate approach to calculations that allow us to investigate the leading order effects of modifications to dispersion relations.

In previous work \cite{davies2023accelerated,tee2022fundamental}, we investigated the effect of modified dispersion relations on accelerated particle detectors and gravitational deflection of light.
In the case of accelerated detectors, we identified that these modifications to the dispersion relations can lead to transition probabilities that are not positive definite when propagation is subluminal.
This result would appear to put into tension the Davies-Fulling-Unruh effect with the existence of a fundamental length scale.
Nevertheless, the dispersion relations themselves are not the full story, as the use of them in isolation causes the model to be in violation of Lorentz invariance.
DSR restores Lorentz invariance by affording momentum space a non-trivial curved geometry.
In our prior calculations we argued around this complication by asserting that any adjustment to the modified propagators we derive must necessarily be small, contributing extra powers in the expansion parameter we used $\kappa\eta^2$, where $\eta$ is the inverse of the Planck Mass, and $\kappa$ a dimensionless parameter controlling the sign of any corrections and intimately linked to whether propagation is subluminal (i.e. for values of $\kappa<0$).
In this work we explore this assertion more thoroughly by taking the calculation back to basics and applying it to the most basic object in quantum field theory, the quantum vacuum.

The effect of momentum curvature enters into the calculation by changing the integration measure when we integrate over momenta.
In the particle detector example described above, we typically need position space propagators which we obtain by a Fourier transform of the momentum space propagators suitably deformed to be consistent with the modified dispersion relations.
Some of the earliest considerations of quantized spacetime dating back to Snyder's original proposal \cite{snyder1947quantized,hossenfelder2013minimal}, modelled Minkowski spacetime as a $3+1$ hyperbolic surface embedded in a higher dimensional ($4+1$) space.
This concept is carried over into many interpretations of DSR, with the additional assumption that momentum space is weakly de Sitter \cite{amelino2016pathways,amelino2012relative}.
As the momentum space now has a non-trivial geometry any integration measure must include a factor of $\sqrt{-g_p}$, where $g_p$ is the determinant of the momentum space metric tensor.

The system we study here is the quantum vacuum of a massless scalar field, both with and without a boundary.
This is the simplest set up in which to test further the calculational framework that exposed the pathologies in the accelerated detector when considering spacetime with a fine-grained structure.
In the absence of a boundary we compute the energy density of a thermalized vacuum, and in the presence of a static boundary we repeat the calculation for the Casimir effect \cite{casimir1948influence}.
In both cases we are computing the energy density of the vacuum, and we do so  using two methods. 
The methods were chosen specifically to rely upon the modified dispersion relations and the non-trivial momentum measure independently.
The first method is the technique of ``point splitting'' to compute a renormalized stress-energy-momentum tensor.
This makes direct use of the propagator of the theory and therefore modified dispersion relations.
The second approach is more direct, computing the energy density from an integral over field modes.
This relies upon the density of states for the thermal vacuum and the Hamiltonian for the Casimir effect, and both involve an integral over momenta.
The two approaches neatly separate the measure from the propagator as the point splitting technique is local and does not require a measure, whereas the integral over modes is global and does not require knowledge of the propagator.
If the two methods agree we can conclude that the momentum measure and modified dispersion relations are internally consistent, which will inform our reevaluation of the results obtained in earlier work.
As we describe in \Cref{sec:framework}, the momentum measure which is consistent with the modified propagators we use here and in previous work, has a first order correction of $\sqrt{-g}=1+\kappa\eta^2p^2$.
The consequence of this for our previous calculation is that to $\BigO{\kappa\eta^2}$ we obtain a similar result, but the sign of the correction will be changed. 
This has the consequence that the tension in the result for the Davies-Fulling-Unruh effect on a fine-grained background is resolved when propagation is subluminal.
This is the first result of this work and confirms that the inclusion of a modified momentum measure restores the Lorentz invariant character to results obtained by using DSR to leading order.

The second set of results in the paper concern the application of these methods to the vacuum energies of free thermal and bounded (Casimir) spacetimes.
When we perform the detailed calculations both methods yield fundamentally consistent results for the energy density of both the free vacuum and the Casimir effect.
For the free vacuum we obtain,
\begin{align}
    \expval{T_{tt}}&=\frac{\pi^2}{30 \beta^4} + \frac{\kappa\eta^2\pi^4}{126 \beta^6} \text{,}\label{eqn:point_split_res}\\
    \expval{E}&=\frac{\pi^2}{30 \beta^4} + \frac{8\kappa\eta^2\pi^4}{126 \beta^6} \text{,} \label{eqn:state_dens_res}
\end{align}
where the first result \Cref{eqn:point_split_res} is obtained using the modified propagator, and the second \Cref{eqn:state_dens_res} using the density of states.
For the second result the straight application of the density of states approach yields a correction that is larger than that from point splitting, but we explore in \Cref{sec:freespace} how this can arise as a result of the over-counting of unphysical states.
We describe in more detail there how by the introduction of a cutoff to avoid over-counting of unphysical states (i.e. wavelengths below the length scale of spacetime) the correspondence can be made exact.
For the Casimir effect \cite{casimir1948influence} we can also obtain the result from modified propagators by the use of `point-slitting' as first used by Brown and Maclay \cite{brown1969vacuum}, to obtain the classical result and a correction term,
\begin{equation}
    \expval{T_{\mu\nu}}=\frac{-\pi^2}{1440a^4}\left ( 1 -\frac{5\kappa\eta^2\pi^2}{42a^2} \dots \right)\mqty[ 
                            1 & 0 & 0 & 0\\
                            0 & -1 & 0 & 0 \\
                            0 & 0 & -1 & 0 \\
                            0 & 0 & 0 & 3 \\
                            ] \text{.}
\end{equation}
Approaching this calculation using the Hamiltonian for the field, and we arrive at an almost identical result, with the exception that the correction term $\frac{12\kappa\eta^2\pi^2}{42a^2}$ is again larger.
In section \Cref{sec:casimir} we explain how this slight difference arises from including non-physical states in the calculation, and how their exclusion could sharpen the correspondence.

The layout of this paper is as follows.
We begin by an overview in \Cref{sec:framework} of the calculation framework in terms of the modifications made to the propagators and the momentum measure.
We turn to the calculation of the free space vacuum energy in \Cref{sec:freespace}, and then the Casimir effect in \Cref{sec:casimir}.
As the computations for these results are lengthy we reserve the details for the appendices to not clutter the arguments, and cover just the major points of the calculations in the text.
At all times we work in units where $\hbar=c=k_B=G=1$, and work in a mostly minus metric signature.

We conclude in \Cref{sec:conclusion}, with a discussion of the results and a survey of future directions.

\section{Modified Propagators and the Curvature of Momentum Space}
\label{sec:framework}
Our approach is to model the presence of a finite scale by the use of modified propagators inspired by DSR to recompute various well-known effects that arise when studying quantum field theory in curved spacetimes.
The modifications extend the canonical energy-momentum dispersion relation to include terms that are higher order in the momentum.
This is the core proscription of both Doubly Special Relativity \cite{amelino2001testable} and also other approaches such as Ho{\u r}ava-Lifshitz theories of gravity (HLG) \cite{hovrava2009quantum}).
We focus on a modified dispersion relation of the form,
\begin{equation}\label{eqn:dispersion_massive}
    E^2=p^2+\kappa \eta^2p^4 +m^2\text{,}
\end{equation}
where $\eta$ is a constant with dimension of inverse mass (usually assumed to be the Planck mass), and $\kappa = \pm 1$ is a dimensionless constant that controls the sign of the modification.
In the case of other modified dispersion relation theories such as HLG, the usual proscription sees higher terms in $p$ such as $p^6$.
The significance of $\kappa$ is that for positive values the equations of motion indicate that propagation can be superluminal, and only for $\kappa < 0$ is the maximum speed of propagation $c$.
It is this ambiguity in the maximum speed of propagation that hints at the non-Lorentzian nature of a theory that uses modified propagators and nothing else.
It is worth noting that the standard dispersion relation $E^2=p^2+m^2$, is a simple consequence of the Minkowski norm of the four vector $p_\mu p^\mu = m^2$ being evaluated in the rest frame of the propagating particle.
One arrives at the standard result assuming that $p_\mu p^\mu =g_p^{\mu\nu}p_\mu p_\nu$, where $g_p^{\mu \nu}$ is the metric of momentum space and is simply $\text{diag}(1,-1,-1,-1)$.

A Lorentz invariant DSR theory implies that momentum space is curved \cite{amelino2012relative}, and as a result the use of modified dispersion relations must be done in a way that respects the non-trivial nature of momentum space.
The norm in a general momentum space is now more complex than the simple $p_\mu p^\mu = m^2$ proscription, and relies upon the arc length of the interval in momentum space along a geodesic.
This intuitively computes the `length' of the momentum four vector in the more general momentum space, in the same way that for non-trivial spacetimes we compute proper time.
The full expression for this norm $D(0,p^\mu)$ is,
\begin{equation}\label{eqn:momentum_norm}
     D(0,p^\mu)=m^2=\int\limits_0^1 \sqrt{g^{\mu\nu}\dot{p}_{\mu}\dot{p}_{\nu}}~\dd s \text{,}
\end{equation}
where we integrate along a geodesic in the momentum space described by a more general interval $ds^2=g_{\mu \nu}\dd p^\mu \dd p^\nu$ (we have dropped the $p$ suffix to denote momentum space metric, and unless otherwise stated metrics are for the momentum space $\mathcal{P}$).
The geodesic curve is parameterized by $s$ and in \Cref{eqn:momentum_norm} the $\dot{p}_{\mu}$ should be interpreted as differentiation with respect to $s$.
We cover the details in \Cref{sec:appendix_c}, but our goal is to obtain a form for $g_{\mu \nu}$ that recovers our modified propagator \Cref{eqn:dispersion_massive} from \Cref{eqn:momentum_norm}.
We discover that the following choice of non-diagonal metric for $n$-dimensional space,
\begin{equation}\label{eqn:metric}
\begin{split}
    &g^{\mu\nu}=\begin{pmatrix}
        1 & 0 \\
        0 & -\cosh^2 \left[\left(\alpha^2\right)^{\frac{1}{2}} p_0 \right] \Omega_{ij} \\
    \end{pmatrix}   \text{~with,~~}\\
    &\Omega_{ij} = \delta_{ij} + \frac{p_ip_j}{\alpha^2-p^2}\text{~~,} i=1,\dots,n \text{,}
\end{split}
\end{equation}
satisfies this constraint.
With this choice of metric, we can then replace the conventional Lorentz covariant momentum measure $\int \dd^4 p = \int \frac{\dd^3 p}{(2\pi)^3E_p}$, where $E_p$ is the on shell energy with $\int \sqrt{-g} \dd^4 p$.
We can expand the metric determinant in powers of $\kappa\eta^2$, and to leading order we obtain,
\begin{equation}\label{eqn:metric_det_leading}
    \sqrt{-g}=1+\kappa\eta^2 p^2 \text{,}
\end{equation}
where $p^2$ here refers to the norm of the spatial momentum (i.e. $p^2=\va{p}^2$), and going forward if we omit a Greek index we assume $p$ is the spatial momentum.

In our prior investigations the importance of this measure is that we require position space Green's functions.
Our starting point is to convert \Cref{eqn:dispersion_massive} into an associated Feynman diagram \cite{tee2022fundamental},
\begin{figure}[H]
\begin{center}
$\vcenter{\hbox{
   \hspace{-0.4\linewidth}
	\raggedright
        \setlength{\parindent}{1.2cm}
        \begin{fmffile}{prules}
        \begin{fmfgraph*}(55,40)
            \fmfleft{o1}
            \fmfright{o2}
            \fmf{dashes_arrow, label=$p$}{o1,o2}
            \fmflabel{}{o1}
            \fmflabel{\Large $=\frac{1}{p_0^2-p^2(1+\kappa \eta^2 p^2)-m^2 + i\epsilon}$,}{o2}
        \end{fmfgraph*}
        \end{fmffile}
        }}$
\end{center}
\end{figure}
and  position space propagator, 
\begin{equation}\label{eqn:lorentz_prop}
    G(t,x;t',x')=\int\limits_{-\infty}^\infty \frac{\dd^n p}{(2\pi)^n} \frac{e^{-i[p_0(t-t') -\va{p}\vdot(\va{x}-\va{x'})]}}{p_0^2-p^2(1+\kappa \eta^2 p^2) - m^2 } \text{.}
\end{equation}
As obtained in \cite{davies2023accelerated} we have for the modified propagator for a massless scalar field in position space, 
\begin{equation}\label{eqn:final31timelike}
    D^+(t,r)=\frac{-\theta(\sigma^2)}{4\pi^2(\sigma^2 - i\epsilon)}\left \{ 1-\frac{\kappa \eta^2}{(\sigma^2-i\epsilon)}\right \} \text{,}
\end{equation}
with the corresponding (retarded) space-like case,
\begin{equation}\label{eqn:final31spacelike}
    D^-(t,r)=\frac{\theta(-\sigma^2)}{4\pi^2(-\sigma^2 + i\epsilon)}\left \{ 1+\frac{\kappa \eta^2}{(-\sigma^2+i\epsilon)}\right \} \text{.}
\end{equation}
For economy of notation $\sigma^2=t^2-r^2$ the invariant interval, with $\theta(x)$ being  the normal sign function.
Setting $\eta=0$ recovers the standard result \cite{hong2010analytic}.

\section{The Free Space Vacuum  in Pixelated Spacetime}
\label{sec:freespace}
\subsection{Energy density from modified propagators}
Our first approach to computing the energy density will make use of the modified propagator described in \Cref{sec:framework}.
The approach exploits the approximate stress energy renormalization technique often referred to as point splitting \cite{birrell1984quantum}.
The details of the calculation are in \Cref{sec:appendix_a}, but our starting point for this calculation is the thermal Greens function for the modified propagator of the massless scalar field in Cartesian coordinates,
\begin{widetext}
\begin{align}
    D^{(1)}_\beta(t,\va{x};t'\va{x}')&=-\frac{1}{2\pi^2}\sum\limits_{m=-\infty}^\infty \left \{ \frac{1}{(t-t'-im\beta)^2-(x-x')^2-(y-y')^2-(z-z')^2}\right \} \label{eqn:thermalprop_standard}\\
    &+\frac{\kappa\eta^2}{2\pi^2}\sum\limits_{m=-\infty}^\infty \left \{ \frac{1}{[(t-t'-im\beta)^2-(x-x')^2-(y-y')^2-(z-z')^2]^2}\right \} \label{eqn:thermalprop_correction}\text{.}
\end{align}
\end{widetext}
It should be understood that the summation will omit the infra-red divergence for $m=0$.

The key to the method is to recognize that the propagator can also be written as an expectation value of a field commutator.
Specifically, $D^{(1)}_\beta(t,\va{x};t'\va{x}') = \ev{\{  \phi(t,\va{x}), \phi(t',\va{x}') \}}{\beta}$, for a vacuum prepared as a mix of states at coolness (inverse temperature) $\beta$, and $D^{(1)}(t,\va{x};t'\va{x}')  = \ev{\{  \phi(t,\va{x}), \phi(t',\va{x}') \}}{0}$ for the zero temperature pure state vacuum.
As the stress energy tensor is generally expressed in terms of derivatives of the field $\phi$, one can convert second derivatives of $\phi$ (and squares of first derivatives) to derivatives of the propagator at two `split' points that we then take the limit of zero separation.

The specific expression for the conformal stress energy tensor is given by \Cref{eqn:energy_momentum}, and we can then exploit the infinite boundary conditions for the field of $\phi = \partial \phi =0$ to simplify the expression to \Cref{eqn:exp_stress}.
As we are dealing with flat space where $g_{\mu\nu}=\text{diag}(1,-1,-1,-1)$ the awkward cross derivatives are not relevant, and we are left with the diagonal entries of the stress tensor to  evaluate,
\begin{equation*}
    \ev{T_{\mu\mu}}{0}=\ev**{(\partial_\mu \phi)^2 -\frac{1}{4}g_{\mu\mu}\partial_\nu \phi ~\partial^\nu \phi}{0} \text{.}
\end{equation*}
Both derivatives are readily computed from the propagator and for the first contribution \Cref{eqn:thermalprop_standard} we obtain the expected standard result  $\expval{T_{tt}}=\frac{\pi^2}{30\beta^4}$.
More care must be taken with the correction \Cref{eqn:thermalprop_correction}, as the additional square in the denominator will involve cross terms such as $(x-x')^2 (y-y')^2$.
Fortunately when the derivatives involved in the point-splitting method are applied, all of these are zero and do not contribute. 
In fact, the only contribution comes from the time derivatives, and so we arrive at a correction term of $\frac{\kappa\eta^2\pi^4}{126\beta^6}$.
We can compute the other diagonal components similarly as described in \Cref{sec:appendix_a}, and we reproduce our final result for the leading order thermal vacuum energy density from \Cref{eqn:point_split_final} here,
\begin{equation}\label{eqn:point_split_thermal}
     \expval{T_{\mu\nu}}=\left ( \frac{\pi^2}{30\beta^4}   + \frac{\kappa\eta^2\pi^4}{126 \beta^6} \right )\mqty[ 
                            1 & 0 & 0 & 0\\
                            0 & \frac{1}{3} & 0 & 0 \\
                            0 & 0 & \frac{1}{3} & 0 \\
                            0 & 0 & 0 & \frac{1}{3} \\
                            ] \text{.}
\end{equation}

\subsection{Energy density from modified measure}
An alternative approach to the calculation involving point splitting proceeds by using a density of states.
From elementary considerations it will be recalled that for modes of angular frequency $\omega$, the density of states $\rho(\omega)$ can be written as,
\begin{equation}
    \rho(\omega)=\frac{\omega^2}{2\pi^2}\dd \omega \text{.}
\end{equation}
The standard derivation computes the number of states enclosed in an octant of a sphere of radius $\omega$ in momentum space, in essence computing the volume of momentum space $\mathcal{P}$ for the system.
In the fundamental length scale scenario we are considering, this must be modified to account for the non-trivial geometry of $\mathcal{P}$.
As discussed in \Cref{sec:framework}, this amounts to inserting a factor of $\sqrt{-g}$ into the integral.
Computing the partition function for our scalar field at coolness $\beta$, allows us to express the energy density as two integrals to $\BigO{\kappa\eta^2}$ that we show in \Cref{eqn:dens_integral}.
We note here that the correction integral $\frac{\kappa\eta^2}{4\pi^2} \int\limits_0^\infty \frac{\omega^5}{e^{\beta\omega}-1}  \dd \omega$ has an infinite upper limit, which of course implies modes of zero wavelength.
Upon evaluation we have our result for the energy density \Cref{eqn:density_final}, that we restate here,
\begin{equation}\label{eqn:density_unrenormalized}
    \expval{E}= \frac{\pi^2}{30 \beta^4}+ \frac{8\kappa\eta^2\pi^4}{126 \beta^6} \text{.}
\end{equation}

We see that the correction is eight times larger than the point splitting result obtained in \Cref{eqn:point_split_thermal}.
A possible cause of this difference is the infinite upper limit in the correction integral.
We are implicitly assuming that spacetime is discrete and so an infinite angular frequency $\omega$ as the upper limit is not sensible and one would expect the integral to over-count the contribution.
If the upper limit did not extend to infinity the integral will have a smaller numerical value.
To progress, we  introduce into the second integral of \Cref{eqn:dens_integral} a cutoff, $\Lambda$ representing the maximum frequency permissible with a wavelength at the scale of the pixelation of spacetime.
This is usually assumed to be the Planck frequency, but at this point is just an arbitrary upper limit.
When we evaluate the integral, we obtain a complex expression involving Polylogarithms (\cite{gradshteyn2014table} \S 9.5 and \S 9.55) $\mathrm{Li}_n (e^{-\beta\Lambda})$ up to sixth order,
\begin{widetext}   
\begin{equation*}
\begin{split}
        &63\beta^6\int\limits_0^\Lambda \frac{\omega^5}{e^{\beta\omega}-1}  \dd \omega =8\pi^6 + 63\beta^5\Lambda^5 \log (1-e^{-\beta \Lambda}) -315\beta^4\Lambda^4\mathrm{Li}_2(e^{-\beta\Lambda})\\
        &- 1260 \beta^3\Lambda^3 \mathrm{Li}_3(e^{-\beta\Lambda}) - 3780 \beta^2\Lambda^2 \mathrm{Li}_4(e^{-\beta\Lambda}) - 7560 \beta\Lambda \mathrm{Li}_5(e^{-\beta\Lambda}) - 7560  \mathrm{Li}_6(e^{-\beta\Lambda}) \text{.}
\end{split}
\end{equation*}
\end{widetext}
This expression can be expanded around $\beta$ for very high temperatures (energies) or small $\beta$ and we obtain a $\Lambda$ dependent approximation,
\begin{equation*}
\begin{split}
    &\int\limits_0^\Lambda \frac{\omega^5}{e^{\beta\omega}-1}  \dd \omega = \\
    &\frac{\pi^6}{63 \beta^6}\left ( \frac{63 \Lambda^5 \beta^5}{5 \pi^6} - \frac{21\Lambda^6 \beta^6}{4 \pi^6} + \frac{3\Lambda^7\beta^7}{4\pi^6} +\BigO{\beta^9} \right )
\end{split}
\end{equation*}
We propose a renormalization or re-scaling argument and choose a $\beta$ dependent value of $\Lambda$, such that the first term in the expansion is simply $1$. 
On such choice would be $\Lambda^5=\frac{5\pi^6}{63}(1+\beta^{-5})$, which would give for our expansion,
\begin{equation*}
    \int\limits_0^\Lambda \frac{\omega^5}{e^{\beta\omega}-1}  \dd \omega = \frac{\pi^6}{63 \beta^6}\left ( 1+\BigO{\beta^n}  \right ) \text{, with $n\geq1$.}
\end{equation*}
Inserting this back into our calculation \Cref{eqn:dens_integral}, we have for our ``renormalized'' energy density,
\begin{equation}\label{eqn:final_vacuum_energy_dens}
    \expval{E}= \frac{\pi^2}{30 \beta^4}+ \frac{\kappa\eta^2\pi^4}{126 \beta^6} + \BigO{\beta}\text{.}
\end{equation}
This is only formally in agreement with the point-splitting results when we take the limit $\beta \rightarrow 0$.
However taking this limit is no longer be consistent with the upper limit being $\Lambda$ we chose as in this limit the energy scale is higher.
This becomes a re-scaling argument similar to renormalization.
In fact, close inspection of the series expansion of the Polylogarithms indicates that the choice of beta dependent $\Lambda$ would have to be modified for each additional power of $\beta$.
We have perhaps motivated if not proven then principle that the excess value of the energy density from the state density calculation is due to an unrealistic infinite limit to the integrals involved.

\section{The Casimir Effect in Pixelated Spacetime}
\label{sec:casimir}

In the previous section our spacetime was free of boundaries, and we were able to obtain consistent results for the energy densities using either a modified propagator or modified momentum measure.
To push the method a little further, we can complicate the calculation by the introduction of boundaries.
The most studied static system that exploits boundary conditions is the Casimir effect \cite{casimir1948influence}, and there are two potential methods to compute the energy density as a function of plate separation.
We follow here the treatment of Brown and Maclay \cite{brown1969vacuum}, to compute the energy density from the field propagator, and then Fierz \cite{fierz1959anziehung} to compute using an integral over momenta.

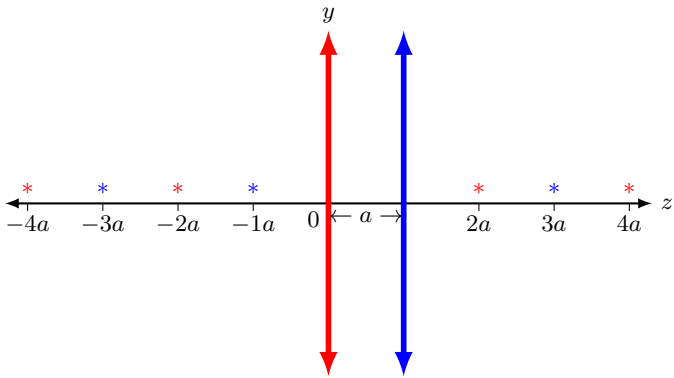
\begin{figure}[hbtp]
    \centering
    \begin{tikzpicture} 
    \def \minx {-4}
    \def \maxx {4}
    \def \miny {-2}
    \def \maxy {2}


    \draw[thick, latex-latex] (\minx-0.3,0)--(\maxx+0.3,0);
    \draw[red,line width=0.75mm,latex-latex] (0,\miny-0.3)--(0,\maxy+0.3);
    \draw[blue,line width=0.75mm,latex-latex] (1,\miny-0.3)--(1,\maxy+0.3);
    \node at (\maxx+0.5,0) {$z$};
    \node at (0,\maxy+0.5) {$y$};

    \foreach \x in {\minx,...,-1}{
        \draw (\x,0)--(\x,-0.1);
        \node[below,black] at (\x,-0.05) {$\x a$};
    }
    \foreach \x in {2,...,\maxx}{
        \draw (\x,0)--(\x,-0.1);
        \node[below,black] at (\x,-0.05) {$\x a$};
    }
    \node[below left,black] at (0,0) {$0$};
    \node[below ,black] at (0.5,0.0) {$\leftarrow a \rightarrow$}; 

    \node[red] at (2,0.2) {$\ast$};
    \node[red] at (4,0.2) {$\ast$};
    \node[red] at (-2,0.2) {$\ast$};
    \node[red] at (-4,0.2) {$\ast$};

    \node[blue] at (3,0.2) {$\ast$};
    \node[blue] at (-1,0.2) {$\ast$};
    \node[blue] at (-3,0.2) {$\ast$};
    
    \end{tikzpicture}
    \caption{For our Casimir system to have a value of $D^{(1)}=0$ on both plates we need an infinite set of image sources placed at a distance of  $z=\pm2an$ for integer $n$ for each plate. Placing the images at a spacing of $a$ locates a source on each plate, and cause double placement of sources at each point along the $z$ axis. This doubling of the boundary terms translates into an extra factor of $2$ on the numerator of the boundary term for the Hadamard function.}
    \label{fig:casimir}
\end{figure}

The general set up is described in \Cref{fig:casimir}, where we have two perfect conductors, infinite in the $x,y$ planes, separated by a small distance $a$.
Although we consider here a massless scalar field, we interpret conducting as enforcing the field to satisfy Dirichlet boundary conditions on the plate $\phi(x,y,0)=\phi(x,y,a)$ for all $x,y$.

We set up the propagator in Cartesian coordinates as a pair of infinite image sums to enforce these boundary conditions, which after subtraction of the free space propagator is,
\begin{widetext}
\begin{equation*}
\begin{split}
    D^{(1)}(t,\va{x};t'\va{x}')&=-\frac{1}{\pi^2}\sum\limits_{n=-\infty}^\infty \left \{ \frac{1}{(t-t')^2-(x-x')^2-(y-y')^2-(z-z'+2an)^2}\right \}  \\
    &+\frac{1}{\pi^2}\sum\limits_{n=-\infty}^\infty \left \{ \frac{\kappa\eta^2}{[(t-t')^2-(x-x')^2-(y-y')^2-(z-z'+2an)^2]^2}\right \} \text{.}
\end{split}
\end{equation*}
\end{widetext}
From here we can follow the same calculation as in the previous section \Cref{sec:freespace}, with the details in \Cref{sec:appendix_b}, and we restate the result \Cref{eqn:casimir_app_splitres} here,
\begin{equation}\label{eqn:casimir_splitres}
    \expval{T_{\mu\nu}}=\frac{-\pi^2}{1440a^4}\left ( 1 -\frac{5\kappa\eta^2\pi^2}{42 a^2} \dots \right)\mqty[ 
                            1 & 0 & 0 & 0\\
                            0 & -1 & 0 & 0 \\
                            0 & 0 & -1 & 0 \\
                            0 & 0 & 0 & 3 \\
                            ] \text{.}
\end{equation}

Turning to an integral over momentum space, the expected energy per unit area \cite{peskin1995introduction} can be expressed as the following divergent integral,
\begin{equation*}\label{eqn:casimir_energy_dens}
    \frac{E(a)}{A}=\frac{1}{2}\sum\limits_{n=1}^\infty\int\limits_{-\infty}^\infty\int\limits_{-\infty}^\infty \frac{\dd p_x \dd p_y}{(2\pi)^2} \sqrt{\left ( \frac{n\pi}{a}\right )+p_x^2+p_y^2} \text{.}
\end{equation*}
The standard approach is to insert a renormalization factor $\exp[ -\alpha E(n)]$, where $E(n)$ is the square root term in the integral and $\alpha$ is a renormalization parameter such that when $\alpha=0$ one recovers the divergent integral.
Our tactic is to insert a factor of $\sqrt{-g}$ term into the measure to account for the non-trivial momentum space geometry.

The details of the calculation are in \Cref{sec:appendix_b}, and we reproduce the result for $\expval{T_{tt}}$ from \Cref{eqn:casimir_app_momres} here,
\begin{equation}\label{eqn:casimir_momres}
    \expval{T_{tt}} = \frac{-\pi^2}{1440 a^4}\left ( 1- \frac{12 \kappa \eta^2 \pi^2}{42 a^2}\right ) \text{.}
\end{equation}
Once again we notice that the correction result is slightly larger than that obtained by point-splitting.
The measure in the integral also suffers from infinite limits, implying modes of infinite momentum, and indeed the sum is also an infinite sum involving modes of zero wavelength.
We speculate that this could be the source of the discrepancy.

The calculation of the renormalized integral is much more complex than the free space version in the prior section, so following the same argument is beyond reach at this point.
We can however explore capping the sum over modes to avoid contributions from modes of wavelength shorter than an arbitrary pixelation length scale.
If we introduce a regulator $\Lambda$ the upper limit of the sum in the correction term \Cref{eqn:correction_sum}, we will have a $\Lambda$ dependent term to subtract from the integral.
A simple computation shows that the integrand would be smaller by,
\begin{equation*}
    \frac{e^{-\frac{\alpha\pi}{a}\Lambda}[e^{\frac{\alpha\pi}{a}}(1+\Lambda)-\Lambda]}{(e^{\frac{\alpha \pi}{a}w}-1)^2} \text{.}
\end{equation*}
As this is always positive (and of course tends to zero as $\Lambda \rightarrow \infty$), we can see that our integral will over count the contribution deriving from the curvature of momentum space associated with the fundamental length scale.

\section{Discussion and Concluding Remarks}
\label{sec:conclusion}
We set out in this work to evaluate the self-consistency of the DSR framework in computing some well-known quantum field theory results.
In our prior work we arrived at the problematic result that for subluminal propagation accelerated detectors do not have a positive definite probability of detecting particles in the vacuum, albeit maintaining a thermal nature to the spectrum.
We remarked that our approach rested solely on the use of propagators modified to agree with the deformed dispersion relations of DSR.
On their own this is not a Lorentz invariant calculation, but requires a modified momentum measure in the production of position space propagators, which can be derived from the metric of momentum space.
This metric, in turn, must be consistent with the modified propagators used.

In section \Cref{sec:framework}, we demonstrated that such a metric exists, and also that to first order the metric would not affect our result for the accelerated detectors.
In essence, the selection of the metric modifies the norm of the momentum four-vector to yield the modified dispersion relations we use in our calculation.
A different modification such as that proposed in HLG would require a different metric.
Taking the simplest quantum vacuum of a free space thermal vacuum, and also one with boundaries we computed the energy density by methods that require one but not both of a modified propagator and momentum measure.
We were able to demonstrate that the results obtained were consistent.
In addition, the form of the modified momentum measure if it were to be used with the modified propagator, would not radically alter the computations involved in the earlier results.
Indeed, for integrals encountered, when involving both the modified propagator and measure to leading order we will be left with an overall $(1+\kappa\eta^2 p^2)^{\frac{1}{2}}$ correction terms.
We can then expand this in the same manner and use it to compute leading order corrections.
What does change is the sign of the contribution, which will alter the values of $\kappa$ that produce non-positive definite transition probabilities.
The natural consequence of this is that the previous Lorentz violating result would hold, but with the pleasing result that propagation below the speed of light produces positive definite probabilities, confirming that the momentum measure restores Lorentz invariance to the calculations.
Put together these results build confidence in the rationality of using the DSR framework in more complex systems to investigate the leading order effects of a fundamental length.
Perhaps more intriguingly, we have a complete set of tools to build upon to analyze quantum field effects in general spacetimes.
As a further example, we could apply these results to curved spacetime examples such as the Einstein universe \cite{ford1976quantum}.
Without providing the details, a simple application of the methods used in this paper (specifically a modified momentum measure), gives the following correction to the energy density $\rho$
\begin{equation*}
    \rho=\frac{1}{480 \pi^2 a^4} -\frac{\kappa\eta^2}{1008 \pi^2 a^6} \text{.}
\end{equation*} 
Here $a$ is the constant scale factor of the Einstein metric $\dd s^2=\dd t^2 - a~ \dd \Omega^2$.
This is of course positive when $\kappa < 0$, that is for subluminal propagation.

Where can we go from here?
As we stated earlier our original interest in the Davies-Fulling-Unruh effect is that it is related to other phenomena such as Hawking radiation.
Although the combination of momentum measure with modified propagator would seem to indicate that the Davies-Fulling-Unruh effect simple acquires a small leading order contribution, it is not self-evident that this is the case for the Hawking effect.
The presence of thermally radiating black-holes is at the origin of the Black-Hole information paradox, and should the pixelation of spacetime cast doubt on that result then the existence of a QG theory would seem to protect us from that paradox.
In ongoing investigations, we intend to test that hypothesis and apply the methods described in this paper and in \cite{davies2023accelerated} to that question.

\appendix
\section{Choice of momentum space metric}
\label{sec:appendix_c}
As has been remarked previously, the use of modified propagators as suggested by DSR on their own, does not yield a theory that preserves Lorentz symmetries.
To restore Lorentz covariance requires other changes to the theory, specifically the admission that momentum space is curved and not trivially Minkowski \cite{amelino2012relative}.
The use of the standard Minkowski metric $g_{\mu \nu}=\text{diag}(1,-1,-1,-1)$, when contracting the four momentum $p^\mu$, yields the normal dispersion relations,
\begin{equation*}
    g_{\mu \nu}p^\mu p^\nu = m^2 = E^2 - \va{p}^2 \text{,}
\end{equation*}
although in what follows when we write powers of $p$ we are referring to the three-momentum $\va{p}$.
This is different from our proposed dispersion relation $E^2=m^2+p^2+ \kappa\eta^2 p^4$, and we ask what choice of momentum space metric is consistent with this form.

To proceed, let us introduce some nomenclature, and we will follow closely the analysis of \cite{amelino2012relative} and private communication with one of the authors \cite{rosati2024private}. 
In this discussion we will denote by $g^{\mu \nu}$ the non-trivial metric of $n$-dimensional momentum space $\mathcal{P}$.
We assume that $\mathcal{P}$ is curved and denote by $C^{\mu \nu}_\rho$, the Christoffel coefficients of the metric $g^{\mu \nu}$ of $\mathcal{P}$.
The norm of the four-momentum is now defined as the geodesic distance from the origin of an an arbitrary point in momentum space $p^\mu=(p^0,p^i)$, where latin indices run from $i=1$ to $i=n-1$.
Using the metric of $\mathcal{P}$ we define the invariant interval $ds$ in the usual way as,
\begin{equation*}
    \dd s^2=g_{\mu\nu}\dd p^\mu \dd p^\nu \text{,}
\end{equation*}
which permits the definition of solutions to the geodesic equation in $\mathcal{P}$,
\begin{equation}\label{eqn:mom_geodesic}
    \ddot{p}_\rho + C^{\mu \nu}_\rho \dot{p}_{\mu}\dot{p}_{\nu} = 0 \text{,}
\end{equation}
where $\dot{p}$ denotes differentiation with respect to $s$.
Once we have a solution to this, our dispersion relation can be read off by computing the arc length along a geodesic from the origin to $p^\mu$, as
\begin{equation}
    D(0,p^\mu)=m^2=\int\limits_0^1 \sqrt{g^{\mu\nu}\dot{p}_{\mu}\dot{p}_{\nu}}~\dd s \text{.}
\end{equation}

A common choice for momentum space metric is to assert that it is de Sitter like \cite{amelino2011principle,amelino2012relative,amelino2016pathways}, with the de Sitter parameter controlling the degree to which the geometry of momentum space $\mathcal{P}$ is curved.
Usually $n$ dimensional de Sitter space is viewed as an $n$-dimensional hyperbolic slice in an $n+1$ dimensional Minkowski space.
For a $4$-dimensional hyperboloid, we can define this as the surface,
\begin{equation}\label{eqn:hyperboloid}
    z_Az^A=z_0^2-z_1^2+z_2^2+z_3^2+z_4^2=-\frac{1}{\alpha^2} \text{,}
\end{equation}
where $A=0,1,2,3,4$ and the Minkowski metric in the $5$ dimensional space $g_{AB}=(+,-,-,-,-)$.
The radius of the hyperboloid $\alpha$ has dimensions of length and in our notation we can associate it to $\eta$.
However for later convenience we actually set $\alpha^2=\frac{2\kappa\eta^2}{n}$, for an $n$-dimensional space.
The general $n$ dimensional hyperboloid has an invariant interval,
\begin{equation}
    \dd s^2=dz_0^2 - \sum\limits_{A=1}^{n} \dd z_A^2 \text{,}
\end{equation}
subject to the constraint $g_{AB}z^Az^B=-\alpha^2$.
This constraint has the effect of reducing the dimensionality of the space by creating a constraint on one of the coordinates, which by convention we choose to be $z_n$ leaving $n$ independent ones.
For conventional use where $n=4$, we have four independent coordinates and constrain $z_4$.

We introduce the following coordinates for $i=1,\dots,n$ (sometimes referred to as a closed slicing),
\begin{align}
    z_0&=\left(\alpha^2\right)^{-\frac{1}{2}}\sinh \left[\left(\alpha^2\right)^{\frac{1}{2}} p_0 \right] \text{,}\\
    z_i&=\cosh \left[\left(\alpha^2\right)^{\frac{1}{2}} p_0 \right] p_i \text{~,with} \\
    &\sum\limits_{i=1}^{i=n} p_i^2 = \left(\alpha^2\right)^{-1}  \text{.} \label{eqn:momconstraint}
\end{align}
These coordinates parameterize the whole of de Sitter space and the $p_i$ coordinates parameterize an $n-1$ dimensional sphere of radius $\alpha^{-2}$, noting the signs in this constraint are positive.
It is easy to check that with the constraint \Cref{eqn:momconstraint} these coordinates satisfy  \Cref{eqn:hyperboloid}, and that in the limit $\eta\rightarrow0$, we recover Minkowski coordinates.
Specifically if we displace the origin of the sphere described by \Cref{eqn:momconstraint} to the point of intersection of the sphere with the $p_n$ axis, i.e  $p_1,\dots,p_{n-1}=0, p_n=\alpha^{-1}$, in the limit $\eta\rightarrow0$ we have
\begin{align*}
    &z_0 \xrightarrow{\eta\rightarrow0} =p_0 \text{,}\\
    &z_1,\dots,z_{i-1} \xrightarrow{\eta\rightarrow 0} =p_i \text{,}\\
    &z_4 \xrightarrow{\eta\rightarrow 0} =0 \text{.}
\end{align*}
In these coordinates our invariant interval now becomes,
\begin{equation}\label{eqn:interval}
    \dd s^2 = \dd p_0^2 -\cosh^2 \left[\left(\alpha^2\right)^{\frac{1}{2}} p_0 \right] \sum\limits_{i=1}^{i=n} \dd p_i^2 \text{,}
\end{equation}
subject to the constraint \Cref{eqn:hyperboloid}, and still involve the $n$ momentum coordinates.
In the limit $\alpha\rightarrow0$, this reduces to the `Minkowski' interval $ds^2=dp_0^2-\sum dp_i^2$.

As an aside, one can further re-parameterize the coordinates by introducing a ``conformal'' time equivalent momentum coordinate,
\begin{equation*}
    \tau=\left(\alpha^2\right)^{-\frac{1}{2}} \tan^{-1} \left[\sinh \left(\alpha^2\right)^{\frac{1}{2}} p_0 \right] \text{,}
\end{equation*}
that illustrates the fact that the interval is conformal to Minkowski as we can write,
\begin{equation}
    \dd s^2 = \cosh^2 \left[\left(\alpha^2\right)^{\frac{1}{2}} p_0 \right]  \left(\dd p_0^2 -\sum\limits_{i=1}^{i=n} \dd p_i^2 \right )\text{.}
\end{equation}
The conformal factor $\cosh^2 \left[\left(\alpha^2\right)^{\frac{1}{2}} p_0 \right]$ can be converted to a function of $\tau$ to give $C(\tau) = \sec^2 \left[\left(\alpha^2\right)^{\frac{1}{2}} p_0 \right]$.
A short calculation, following \cite{birrell1984quantum}, can then compute from the conformal factor the scalar curvature of the momentum space as,
\begin{equation}
    R=6 \left(\alpha^2\right) \left \{ 1+\sin^2 \left[ \left(\alpha^2\right)^{\frac{1}{2}} p_0\right] \right \}\text{.}
\end{equation}
This result confirms our intuition that as our scale parameter $\alpha\rightarrow 0$, the scalar curvature of the momentum space is zero, consistent with it being Minkowski in this limit.

However, we wish to further constrain the embedding and eliminate the $p_n$ coordinate, using the constraint \Cref{eqn:momconstraint}, re-written as,
\begin{equation*}
    p_n=\sqrt{\alpha^2 - \sum\limits_{i=1}^{n-1} p_i^2} = \sqrt{\alpha^2-p^2} \text{.}
\end{equation*}
This has the consequence of creating a non-diagonal $n-1$ dimensional metric because when we differentiate the constraint we find,
\begin{equation*}
    \dd p_n^2=\frac{1}{\alpha^2-p^2}\sum\limits_{i,j}^{n-1} p_ip_j\dd p_i \dd p_j \text{.}
\end{equation*}
After elimination of $p_n$ we have the following $n$ dimensional metric on Cartesian momentum coordinates,
\begin{equation}\label{eqn:metric_mom}
\begin{split}
    &g^{\mu\nu}=\begin{pmatrix}
        1 & 0 \\
        0 & -\cosh^2 \left[\left(\alpha^2\right)^{\frac{1}{2}} p_0 \right] \Omega_{ij} \\
    \end{pmatrix}   \text{,~~}\\
    &g_{\mu\nu}=\begin{pmatrix}
        1 & 0 \\
        0 & -\sech^{2} \left[\left(\alpha^2\right)^{\frac{1}{2}} p_0 \right] \Omega^{-1}_{ij}\\ \text{.}
        \end{pmatrix}
\end{split}
\end{equation}
The cross terms $\Omega_{ij}$,$\Omega^{-1}_{ij}$ are defined as,
\begin{equation}
    \Omega_{ij} = \delta_{ij} + \frac{p_ip_j}{\alpha^2-p^2} \text{~,~} \Omega^{-1}_{ij} = 1 -\alpha^2p_i p_j \text{.}
\end{equation}
In the above it should be understood that \Cref{eqn:metric_mom} describes an  $n\times n$ metric, and for the $\Omega_{ij}$ terms $i,j=1,\dots,n-1$.
To obtain the leading order correction to the on-shell relation and metric determinant we can expand \Cref{eqn:metric_mom} to $\BigO(\kappa\eta^2)$, to obtain,
\begin{equation}\label{eqn:metric_leading}
\begin{split}
    &g^{\mu\nu}=\begin{pmatrix}
        \\[-2mm]
        1 & 0 \\
        0 & -\delta_{ij}-\alpha^2[p_0^2\delta_{ij}+p_i p_j] \\
        \\[-2mm]
    \end{pmatrix}   \text{,~~}\\
    &g_{\mu\nu}=\begin{pmatrix}
        \\[-2mm]
        1 & 0 \\
        0 & -\delta_{ij}+\alpha^2(p_0^2\delta_{ij}+p_i p_j)\\
        \\[-2mm]
        \end{pmatrix} \text{.}
\end{split}
\end{equation}

We compute the momentum space Christoffel symbols using the equivalent of the position space formula, 
\begin{equation*}
    C^{\mu \nu}_\rho = \frac{1}{2}g_{\rho\sigma}(g^{\sigma\mu,\nu}+g^{\nu\sigma,\mu}-g^{\mu\nu,\sigma}) \text{.}
\end{equation*}
Applying this to \Cref{eqn:metric_leading} we find the only non-zero coefficients are, 
\begin{equation}\label{eqn:chris_leading}
\begin{split}
    &C^{jk}_0=C^{0k}_j=C^{k0}_j=\alpha^2p_0\delta_{jk} \text{~,~} \\
    &C^{kl}_j=\alpha^2p_j \delta_{ij} \text{,}\\
    &C^{0jk}=C^{j0k}=-C^{jk0}=-\alpha^2p_0 \delta_{jk} \text{~,~}\\
    &C^{jkl}=-\alpha^2\delta_{jk}p_l \text{.}
\end{split}
\end{equation}

The geodesic equation can be solved, and following the analysis in \cite{amelino2012relative}, our solution for $D(0,p^{\mu})$ is,
\begin{equation*}
    D(0,p^{\mu})=m^2=p_\mu p^\mu + C^{\rho\mu\nu}p_\rho p_\mu p_\nu \text{.}
\end{equation*}
Inserting our leading order Christoffel symbols \Cref{eqn:chris_leading}, we have,
\begin{equation*}
    m^2=p_0^2-p^2-\alpha^2(p_0^2+p^2)p^2 \text{.}
\end{equation*}
For a massless particle, and noting a zero order we have $0=p_0^2-p^2$, and that $E=p_0$, we can for the case of $n=4$, recover the on-shell dispersion relations we have used in the text,
\begin{equation*}
    E^2=p^2 + \kappa\eta^2 p^4 \text{.}
\end{equation*}
We conclude that our proposed metric \Cref{eqn:metric_mom} is consistent with the modified dispersion relations used in the analysis in the form of our  modified propagators.

Turning to the integral measure, we can compute the leading order correction to $\sqrt{-g}$ using \Cref{eqn:metric_mom} that we will use to perform our density of states computations.
We essentially have two approaches we can follow to compute this, either by expanding directly expanding the determinant of \Cref{eqn:metric_mom}, such that,
\begin{equation*}
    g=\det( g^{\mu\nu})=-\frac{\cosh^{2(n-1)}(\alpha^2p_0)}{1-\alpha^2p^2} \text{,}
\end{equation*}
or by exploiting the relationship between the determinant and the trace of a matrix that is close to the identity.
Expanding directly this expression for the determinant, and taking the square root we obtain,
\begin{equation*}
    \sqrt{-g} = 1+\frac{\alpha^2}{2}((n-1)p_0^2 +p^2) = 1+\frac{n\alpha^2}{2} p_0^2 \text{,}
\end{equation*}
for massless particles.

We can exploit the relationship between the trace and the determinant of an $n\times n$ matrix $\mathbf{M}_n$ near the identity given by,
\begin{equation}\label{eqn:det_trace}
    det(\mathbf{I}_n + h\mathbf{M}_n) = 1+ h\Tr(\mathbf{M}_n) + \BigO{h^2} \text{,}
\end{equation}
for small h, and the fact that $det(k\mathbf{M}) = k^n det(\mathbf{M})$ to compute $\sqrt{-g}$.
The determinant, only depends upon the lower spatial part as it is diagonal in the $p_0$ component.
For the spatial part of the metric we can write this as,
\begin{equation}
    g^{\mu\nu} = -1(\mathbf{1} + \alpha^2 B^{\mu\nu}) \text{,}
\end{equation}
where $\mathbf{1}_n$ is the $n\times n$ unit matrix and $B^{\mu\nu}$ is the matrix,
\begin{equation}
    B^{\mu\nu}=\begin{pmatrix}
        (p_0^2+p_1^2) & p_1p_2 & p_1p_3  \\
        p_1p_2 & (p_0^2+p_2^2) & p_2p_3  \\
        p_1p_3 & p_2p_3 & (p_0^2+p_3^2)  \\
    \end{pmatrix} \text{.}
\end{equation}
Inserting this definition into \Cref{eqn:det_trace} we have,
\begin{equation}
    \sqrt{-g} = \left[1+\alpha^2\{(n-1)p_0^2 + p^2\}\right]^{\frac{1}{2}} \text{.}
\end{equation}
For a massless particle to zeroth order $p^2=p_0^2$ (higher order on-shell terms will not contribute to this leading order estimate), and expanding the square root we have,
\begin{equation}
    \sqrt{-g}=1+\frac{n\alpha^2}{2} p_0^2 \text{,}
\end{equation}
in agreement with the result obtained by directly computing the determinant of \Cref{eqn:metric_mom}.
Finally, setting $n=4$, and inserting our definition of $\alpha$, we arrive at our final result,
\begin{equation}\label{eqn:first_order_measure}
    \sqrt{-g} = 1+\kappa \eta^2 p_0^2 \text{~.}
\end{equation}
This approximate measure will be used in the text to compute energy densities from state densities, and is fully consistent with our on-shell dispersion relations to order $\mathcal{O}(\kappa\eta^2)$.

\section{Calculating the Free Vacuum Energy Density}
\label{sec:appendix_a}
\subsection{Point splitting and the modified propagator}
\label{sec:point_split_app}
Our starting point is the free space propagator for a massless scalar field with correction term, from \Cref{eqn:final31spacelike,eqn:final31timelike}, giving us for the Hadamard function,
\begin{equation}\label{eqn:hadamard_fun}
    D^{(1)}(\sigma^2)=D^+(\sigma^2)+D^-(\sigma^2)= -\frac{1}{2\pi^2} \left \{ \frac{1}{\sigma^2} - \frac{\kappa\eta^2}{\sigma^4}\right \} \text{,}
\end{equation}
where $\sigma^2=\Delta t^2 -\Delta \va{x}^2 -i\epsilon$ and $\sigma^4=\Delta t^4 -\Delta \va{x}^4 -i\epsilon$.
Technically, in Cartesian coordinates the $\sigma^4$ terms contains awkward cross terms $\Delta t^2 \Delta \va{x}^2$, but elementary calculation shows these do not contribute to the result.
To compute the thermal Green's function $D^{(1)}_\beta$, we follow the normal procedure to replace $\Delta t \rightarrow \Delta t -im\beta$, and then sum over all $m$, i.e.  $D^{(1)}_\beta=\sum\limits_{m=-\infty}^\infty D^{(1)}(\Delta t-im\beta,\Delta \va{x})$, with $\beta=(k_B T)^{-1}$.

For the standard propagator in Cartesian coordinates we have,
\begin{widetext}
\begin{equation}\label{eqn:standard_cart_prop}
    D^{(1)}_\beta(t,\va{x};t'\va{x}')=-\frac{1}{2\pi^2}\sum\limits_{m=-\infty}^\infty \left \{ \frac{1}{(t-t'-im\beta)^2-(x-x')^2-(y-y')^2-(z-z')^2}\right \} \text{.}
\end{equation}
\end{widetext}

For a massless scalar field $\phi$, in conformal spacetime we have the following expression for the stress-energy-momentum tensor $T_{\mu\nu}$ \cite{birrell1984quantum},
\begin{equation}\label{eqn:energy_momentum}
    T_{\mu\nu}=\frac{2}{3}\partial_\mu \phi \partial_\nu \phi -\frac{1}{6}g_{\mu\nu}g^{\rho\sigma}\partial_\rho\phi\partial_\sigma\phi - \frac{1}{3}\phi \partial_\mu \partial_\nu\phi +\frac{1}{12}\phi \square \phi \text{,}
\end{equation}
with $\xi=1/6$.
Assuming flat space, we set $g_{\mu\nu}=\text{diag}(1,-1,-1,-1)$.
We seek to compute $\ev{T_{\mu\nu}}{0}$, which simplifies \Cref{eqn:energy_momentum} significantly as this involved integrating over all space.
We note the following identities,
\begin{align*}
    \partial_\mu( \phi\partial_\nu \phi)&=2(\partial_\mu \phi \partial_\nu \phi+\phi \partial_\mu \partial_\nu\phi) \text{, and}\\
    \partial_\mu(\phi \partial^\mu \phi)&=2(\partial_\mu \phi \partial^\mu \phi +\phi \square \phi) \text{.}\\
\end{align*}
When we take the integral, the left-hand side of both is a total derivative, and so by Gauss's theorem, and the fact that we assume at infinity that $\partial \phi=\phi=0$, we have,
\begin{align*}
    \int_{\mathcal{M}} \partial_\mu( \phi\partial_\nu \phi) ~\dd^4 x&= \int_{\mathcal{\partial M}} \phi\partial_\nu \phi ~\dd^4 x= 0 \\
    \int_{\mathcal{M}} \partial_\mu( \phi\partial^\mu \phi) ~\dd^4 x &= \int_{\mathcal{\partial M}} \phi\partial^\mu \phi ~\dd^4 x = 0 \text{.}
\end{align*}
Further as we are only considering diagonal entries for the stress-energy-momentum tensor, we have the effective value for the expectation value,
\begin{equation}\label{eqn:exp_stress}
    \ev{T_{\mu\nu}}{0}=\ev**{\left(\partial_\mu \phi \partial_\nu \phi -\frac{1}{4}g_{\mu\nu}\partial_\mu\phi \partial^\mu \phi\right)}{0} \text{.}
\end{equation}

From here we note that as $D^{(1)}(x,x')=\ev{\{\phi(x)\phi(x')\}}{0}$, we can state the following identity,
\begin{equation}
    \ev{(\partial_x \phi)^2}{0}=\ev**{\lim_{x'\rightarrow x} \partial_x \phi \partial_{x'} \phi}{0} = \frac{1}{2}\lim_{x'\rightarrow x}\partial_x\partial_{x'} D^{(1)}(x,x') \text{.}
\end{equation}
For the standard thermal propagator including the leading $-\frac{1}{2\pi^2}$, and intentionally not simplifying the fractions we have,
\begin{align*}
    &\lim_{t'\rightarrow t}\partial_t\partial_{t'} D^{(1)}_\beta=\frac{6}{2\pi^2}\sum\limits_{-\infty}^\infty \frac{1}{(\beta m)^4} \\
    &\begin{rcases}
    &\lim_{x'\rightarrow x}\partial_x\partial_{x'} D^{(1)}_\beta \\
    &\lim_{y'\rightarrow y}\partial_y\partial_{y'} D^{(1)}_\beta \\
    &\lim_{z'\rightarrow z}\partial_z\partial_{z'} D^{(1)}_\beta
    \end{rcases}
    =\frac{2}{2\pi^2}\sum\limits_{-\infty}^\infty \frac{1}{(\beta m)^4} \text{.}
\end{align*}
We can now compute the diagonal components of the stress-energy-momentum tensor (all other terms are zero), by inserting these into \Cref{eqn:exp_stress}. 
For example, for the energy density we have $T_{tt}=\frac{3}{4}(\partial_t \phi)^2+\frac{1}{4}\{(\partial_x \phi)^2 +(\partial_y \phi)^2 + (\partial_z \phi)^2\}$.
Omitting the infinite term at $m=0$, we obtain for our first term,
\begin{equation}
    \expval{T_{tt}}= \frac{6}{2\pi^2} \sum\limits_1^\infty \frac{1}{(\beta m)^4}= \frac{\pi^2}{30 \beta^4} \text{.}
\end{equation}
The remaining components are computed simply and each have the value of $\frac{\pi^2}{90\beta^4}$.

For the correction term we have the following propagator, 
\begin{widetext}
\begin{equation}\label{eqn:standard_cart_prop}
    D^{'(1)}_\beta(t,\va{x};t'\va{x}')=\frac{\kappa\eta^2}{2\pi^2}\sum\limits_{m=-\infty}^\infty \left \{ \frac{1}{[(t-t'-im\beta)^2-(x-x')^2-(y-y')^2-(z-z')^2]^2}\right \} \text{.}
\end{equation}
\end{widetext}
As indicated earlier the extra two powers in the denominator lead to awkward cross terms, but the point splitting derivatives simplify matters. 
Computation of the derivatives demonstrate that all terms involving a spatial term gives a value of $\lim_{x'\rightarrow x}\partial_x\partial_{x'} D^{'(1)}_\beta=0$.
Accordingly, we have,
\begin{align*}
    &\lim_{t'\rightarrow t}\partial_t\partial_{t'} D^{(1)}_\beta=\frac{20}{2\pi^2}\sum\limits_{-\infty}^\infty \frac{1}{(\beta m)^6} \\
    &\begin{rcases}
    &\lim_{x'\rightarrow x}\partial_x\partial_{x'} D^{(1)}_\beta \\
    &\lim_{y'\rightarrow y}\partial_y\partial_{y'} D^{(1)}_\beta \\
    &\lim_{z'\rightarrow z}\partial_z\partial_{z'} D^{(1)}_\beta
    \end{rcases}
    =0 \text{.}
\end{align*}
Again we can use \Cref{eqn:exp_stress} to compute our diagonal components, and again we perform the sum ommiting the $m=0$ divergent term.
Bringing this all together we arrive at our final result,
\begin{equation}\label{eqn:point_split_final}
    \expval{T_{\mu\nu}}=\left ( \frac{\pi^2}{30\beta^4}   + \frac{\kappa\eta^2\pi^4}{126 \beta^6} \right )\mqty[ 
                            1 & 0 & 0 & 0\\
                            0 & \frac{1}{3} & 0 & 0 \\
                            0 & 0 & \frac{1}{3} & 0 \\
                            0 & 0 & 0 & \frac{1}{3} \\
                            ] \text{.}
\end{equation}

\subsection{Density of states with modified momentum measure}
If we assume that the vacuum state of the scalar massless field has the energy spectrum, $E=n\omega$, with $\omega$ the angular frequency of the mode (recall $\hbar=1$) and $n=0,1,2,3,\dots$, one can immediately write the partition function for the canonical ensemble at inverse temperature $\beta$ as,
\begin{equation}\label{eqn:partition_func}
    Z=\sum\limits_{n=0}^\infty e^{-\beta n \omega} = \frac{1}{1-e^{-\beta\omega}} \text{.}
\end{equation}
The standard textbook calculation uses the thermodynamic relation $\expval{E}=-\pdv{\log Z}{\beta}$, to compute the energy density by combining \Cref{eqn:partition_func} with a density of states and integrating over all possible $\omega$.
However, this approach is inconsistent with the assumption of minimal length, as the upper limit of the integral is $\omega \rightarrow \infty$ that implies a zero wavelength.
To accommodate an upper limit of $\omega$, it would seem necessary to introduce a cut off.

In DSR the introduction of a minimal length, whilst maintaining Lorentz covariance, is done by assuming that momentum space is not flat, as described in \Cref{sec:appendix_c}.
This minimal length performs a similar function to the upper momentum cut off by imposing a minimum wavelength.
We seek to compare our point splitting energy density with a calculation using a choice of metric that is consistent with the modified dispersion relations.
We analyzed this in \Cref{sec:appendix_c}, and inserting $\omega$ as our time-like component we have to leading order for our momentum space metric determinant,
\begin{equation}
    \sqrt{-g} = 1+\kappa \eta^2 \omega^2 \text{~.}
\end{equation}

For a  covariant momentum measure, we need to replace $\int \dd^4 p$ with $\int \sqrt{-g_p}\dd^4 p$.
As discussed in \Cref{sec:freespace}, the density of states is computed using the volume of momentum space and so should include the $\sqrt{-g}$ term in the integral measure.
Accordingly the modified density of states is,
\begin{equation}\label{eqn:dens_states}
    \rho(\omega)=\frac{\omega^2 \sqrt{-g}\dd \omega}{2 \pi^2} \text{,}
\end{equation}
which gives for $\log Z$ on substitution,
\begin{equation}\label{eqn:logZ}
    \log Z = \frac{1}{2\pi^2}\int\limits_0^\infty \omega^2 \log Z \sqrt{-g} \dd \omega \text{.}
\end{equation}
From here we expand the exponential from the metric determinant and we arrive at two integrals at order $\BigO{\kappa\eta^2}$,
\begin{equation}\label{eqn:dens_integral}
\begin{split}
    \expval{E}&=-\pdv{\log Z}{\beta} \\
    &=\frac{1}{2\pi^2}\left \{ \int\limits_0^\infty \frac{\omega^3}{e^{\beta\omega}-1}  \dd \omega + \kappa\eta^2 \int\limits_0^\infty \frac{\omega^5}{e^{\beta\omega}-1}  \dd \omega\right \}
\end{split}
\end{equation}
These integrals are elementary, with the first evaluating to $\frac{\pi^4}{15 \beta^4}$ and the second to $\frac{8\pi^6}{63 \beta^6}$.
Substituting in we arrive at our final result,
\begin{equation}\label{eqn:density_final}
    \expval{E}= \frac{\pi^2}{30 \beta^4}+ \frac{8\kappa\eta^2\pi^4}{126 \beta^6} \text{,}
\end{equation}
which is similar to \Cref{eqn:point_split_final}, although larger by a factor of four.
We discuss in \Cref{sec:freespace} how we can account for this difference from the over-counting of unphysical states.

We can also compute the partition function by simply integrating the density of states with the modified measure given by \Cref{eqn:dens_states} to obtain,
\begin{equation}
    Z=\frac{\zeta(3)}{\pi^2 \beta^3}+\frac{12 \kappa \eta^2\zeta(5)}{\pi^2\beta^5} \text{,}
\end{equation}
where $\zeta(n)$ is the Riemann zeta function.
For completeness we can also use the relation $S=\pdv{T \log Z}{T}$ (in units where $k_B=1$), to compute the entropy density, which when re-expressed in terms of a derivative in $\beta$ is,
\begin{equation*}
    S=-\beta^2 \pdv{\beta} \left ( \frac{\log Z}{\beta}\right) \text{.}
\end{equation*}
Using \Cref{eqn:logZ}, and expanding in $\kappa \eta^2$, we have two integrals to compute,
\begin{equation*}
\begin{split}
    \expval{S}&=-\pdv{\beta \log Z}{\beta}=\\
    &\frac{1}{2\pi^2} \int\limits_0^\infty \left \{\frac{\beta\omega^3}{e^{\beta\omega}-1}+\omega^2\log[1-e^{-\beta\omega}] \right\}  \dd \omega \\
    &+ \frac{\kappa\eta^2}{2\pi^2} \int\limits_0^\infty \left \{\frac{\beta\omega^5}{e^{\beta\omega}-1}+\omega^4\log[1-e^{-\beta\omega}] \right\}  \dd \omega \text{.}
\end{split}
\end{equation*}
These may be evaluated by elementary means and give us for the entropy density,
\begin{equation}\label{eqn:entropy_dens}
    \expval{S}=\frac{2\pi^2}{45 \beta^3} + \frac{16\kappa\eta^2\pi^4}{315\beta^5} \text{.}
\end{equation}

\section{Computing the Casimir Effect by Point Splitting}
\label{sec:appendix_b}
\subsection{Casimir effect from point splitting}
We follow the treatments of $\S4.3$ of \cite{birrell1984quantum}, and also \cite{brown1969vacuum} to compute the energy momentum tensor and thus energy density associated with the vacuum of a scalar field in the presence of two boundaries.
The fundamental approach is the same as in \Cref{sec:appendix_a}, where we obtain an expression for the Green's function for the setup of two perfectly conducting plates separated by a distance $a$, and then use the technique of point splitting to obtain an approximate value for the expectation value of the stress-energy-momentum tensor.
The complication arises from the presence of two boundaries that due to their perfectly conducting nature impose a Dirichlet boundary condition on the Green's function.
We sketch the set-up in \Cref{fig:casimir}, and use the method of images to obtain the full propagator $D^{(1)}$ for the scalar field in the presence of this boundary.
We note that to ensure the boundary condition is met on both plates, we need to infinite series of images, one for each plate.
As the series of images stretches to infinity, we can write one image sum, with an additional factor of $2$.

Using our modified propagator \Cref{eqn:hadamard_fun}, the full expression for the propagator including the image term is,
\begin{widetext}
\begin{equation}
\begin{split}
    D^{(1)}(t,\va{x};t'\va{x}')&=D^{(1)}_0(t,\va{x};t'\va{x}')-\frac{1}{\pi^2}\sum\limits_{n=-\infty}^\infty \left \{ \frac{1}{(t-t')^2-(x-x')^2-(y-y')^2-(z-z'+2an)^2}\right \}  \\
    &+\frac{1}{\pi^2}\sum\limits_{n=-\infty}^\infty \left \{ \frac{\kappa\eta^2}{[(t-t')^2-(x-x')^2-(y-y')^2-(z-z'+2an)^2]^2}\right \} \text{.}
\end{split}
\end{equation}
\end{widetext}
To compute the expectation value in this scenario, we subtract away the free space propagator $D^{(1)}_0(t,\va{x};t'\va{x}')$, and utilize the point splitting technique to relate terms in derivatives of the fields to propagators such as,
\begin{equation}
\begin{split}
    \ev{(\partial_\mu \phi)^2}{0}&=\ev**{\lim_{x_\mu'\rightarrow x_\mu} \partial_{x_\mu} \phi ~\partial_{x_\mu'} \phi}{0}\\
    &= \frac{1}{2}\lim_{x_\mu'\rightarrow x_\mu} \partial_{x_\mu}\partial_{x_\mu'} D^{(1)}(x_\mu,x_\mu') \text{.}
\end{split}
\end{equation}
We can then  evaluate the expectation of the appropriate stress-energy-momentum tensor, and in this treatment, we use the ``new improved stress tensor" \Cref{eqn:energy_momentum}.

Dealing with the first (unmodified) term in the propagator, we note that \Cref{eqn:energy_momentum} reduces to \Cref{eqn:exp_stress}, and so we require the relevant derivatives of the propagator, which are as follows,
\begin{align*}
    &~\lim_{t'\rightarrow t}\partial_t\partial_{t'} D^{(1)}=\frac{2}{\pi^2}\sum\limits_{-\infty}^\infty \frac{1}{(2an)^4} \\
    &\begin{rcases}
    &\lim_{x'\rightarrow x}\partial_x\partial_{x'} D^{(1)} \\
    &\lim_{y'\rightarrow y}\partial_y\partial_{y'} D^{(1)}
    \end{rcases}
    =-\frac{2}{\pi^2}\sum\limits_{-\infty}^\infty \frac{1}{(2an)^4} \\
    &~\lim_{z'\rightarrow z}\partial_z\partial_{z'} D^{(1)} =-\frac{6}{\pi^2}\sum\limits_{-\infty}^\infty \frac{1}{(2an)^4} \text{.}
\end{align*}
For the infinite summations on the right-hand side, we omit the infinite value at $n=0$, and note the standard result that $\sum\limits_{-\infty}^\infty \frac{1}{(2an)^4} = \frac{\pi^4}{720 a^4}$.
Substituting into \Cref{eqn:exp_stress} we arrive at the well known ``book value'' for $\expval{T_{\mu\nu}}$,
\begin{equation*}
    \expval{T_{\mu\nu}}=\frac{-\pi^2}{1440a^4}\mqty[ 
                            1 & 0 & 0 & 0\\
                            0 & -1 & 0 & 0 \\
                            0 & 0 & -1 & 0 \\
                            0 & 0 & 0 & 3 \\
                            ] \text{.}
\end{equation*}
Turning to the correction term, the calculation follows the same path as above, but we have the extra two powers in the denominator to deal with.
However, the extra powers actually simplify matters, as any term involving $(t-t')^2,(x-x')^2$ or $(y-y')^2$ are zero.
This simplifies the computation as the only surviving propagator term that contributes to \Cref{eqn:exp_stress} is,
\begin{equation*}
    \lim_{z'\rightarrow z}\partial_z\partial_{z'} D^{(1)}=-\frac{20\kappa\eta^2}{\pi^2}\sum\limits_{-\infty}^\infty \frac{1}{(2an)^6} \text{.}
\end{equation*}
Again, we perform the sum omitting the divergence at $n=0$, which gives the result $\sum\limits_{-\infty}^\infty \frac{1}{(2an)^6}=\frac{\pi^6}{30240 a^6}$.
We can then substitute into \Cref{eqn:exp_stress} to obtain for the correction term,
\begin{equation*}
    \expval{T_{\mu\nu}}=\frac{\kappa\eta^2\pi^4}{12096a^6}\mqty[ 
                            1 & 0 & 0 & 0\\
                            0 & -1 & 0 & 0 \\
                            0 & 0 & -1 & 0 \\
                            0 & 0 & 0 & 3 \\
                            ] \text{.}
\end{equation*}
Bringing this all together we arrive at our final result,
\begin{equation}\label{eqn:casimir_app_splitres}
    \expval{T_{\mu\nu}}=\frac{-\pi^2}{1440a^4}\left ( 1 -\frac{5\kappa\eta^2\pi^2}{42 a^2} \dots \right)\mqty[ 
                            1 & 0 & 0 & 0\\
                            0 & -1 & 0 & 0 \\
                            0 & 0 & -1 & 0 \\
                            0 & 0 & 0 & 3 \\
                            ] \text{.}
\end{equation}

\subsection{Casimir effect from energy density}
\label{sec:casimir_density}
When analyzing the free space vacuum, we calculated the energy density from both point splitting and density of states.
With the Casimir effect our alternative approach utilizing the modified momentum measure explored in \Cref{sec:appendix_c} starts with the regularized energy density.
The treatment follows the original calculation by Fierz \cite{fierz1959anziehung}, which uses as a starting point the energy per unit area of a massless scalar field constrained by two infinite area plates separated by a distance $a$.
Using the same set up in \Cref{fig:casimir}, we can write the energy per unit area as,
\begin{widetext}
\begin{equation}\label{eqn:casimir_energy_dens}
    \frac{E(a,\alpha)}{A}=\frac{1}{2}\sum\limits_{n=1}^\infty\int\limits_{-\infty}^\infty\int\limits_{-\infty}^\infty \frac{\dd p_x \dd p_y}{(2\pi)^2} \sqrt{\left ( \frac{n\pi}{a}\right )+p_x^2+p_y^2}~\exp\left\{ -\alpha \sqrt{\left ( \frac{n\pi}{a}\right )+p_x^2+p_y^2} \right\} \text{.}
\end{equation}
\end{widetext}
The energy is formally divergent and the exponential function is introduced to regularize the integral.
Setting $\alpha=0$ recovers the divergent energy per unit area, and to arrive at the energy per unit volume $\expval{T_{tt}}$ we will divide our result by $a$.

The introduction of the momentum measure proceeds by introducing the metric determinant as a first order approximation from \Cref{eqn:first_order_measure}.
For a massless scalar field, we note that $p_0^2-\sum\limits_i p_i^2=0$, and that the terms inside the square root in \Cref{eqn:casimir_energy_dens} are simply $\sum\limits_i p_i^2$, and so we can re-write the momentum measure as,
\begin{equation*}
    \sqrt{-g}=1+\kappa\eta^2 \left [ \left ( \frac{n\pi}{a}\right )+p_x^2+p_y^2 \right ] \text{.}
\end{equation*}
We can now rewrite the double integrals in polar coordinates, and make a substitution by defining $z=(\frac{a}{n\pi})^2[p_x^2+p_y^2]$, such that with $r^2=p_x^2+p_y^2$, $r \dd r = \frac{1}{2}(\frac{n\pi}{a})^2 \dd z$.
With the modified measure we are left with two integrals,
\begin{widetext}
\begin{equation*}
    \frac{E(a,\alpha)}{A}=\frac{1}{8\pi} \sum\limits_1^\infty \left (\frac{n\pi}{a} \right)^3 \int\limits_0^\infty \sqrt{1+z} \exp \left \{ -\alpha \frac{n\pi}{a}\sqrt{1+z} \right \} ~ \dd z +\frac{\kappa \eta^2}{8\pi} \sum\limits_1^\infty \left (\frac{n\pi}{a} \right)^5 \int\limits_0^\infty (1+z)^{3/2} \exp \left \{ -\alpha \frac{n\pi}{a}\sqrt{1+z} \right \} ~ \dd z
\end{equation*}
\end{widetext}
A further substitution of $w=\sqrt{1+z}$ can be made, and we note that by differentiating with respect to $\alpha$, we can further simplify to,
\begin{align}
    \frac{E(a,\alpha)}{A}&=-\frac{1}{4\pi} \sum\limits_1^\infty \dv[3]{\alpha} \int\limits_0^\infty \frac{\exp[-\alpha (\frac{n\pi}{a})w]}{w} ~ \dd w  \\
    &-\frac{\kappa \eta^2}{4\pi} \sum\limits_1^\infty \dv[5]{\alpha} \int\limits_0^\infty \frac{\exp[-\alpha (\frac{n\pi}{a})w]}{w} ~ \dd w \label{eqn:correction_sum}
\end{align}
In both of these integrals we can perform one differentiation by $\alpha$, which then presents an elementary summation, which leaves the integrand in both cases to be in the following form,
\begin{equation*}
    \int\limits_0^\infty \frac{e^{\frac{\alpha \pi}{a}w}}{(e^{\frac{\alpha \pi}{a}w}-1)^2} ~ \dd w = \left (\frac{\alpha\pi}{a}\right)^{-1} \frac{1}{e^{\frac{\alpha \pi}{a}}-1}
\end{equation*}
We can now exploit the exponential generating identity for the Bernoulli numbers (\cite{gradshteyn2014table} \S 9.62), 
\begin{equation*}
    \frac{x}{e^x-1}=\sum\limits_{m=0}^\infty \frac{B_m}{m!}x^m \text{,}
\end{equation*}
to arrive at the following result for the integrals,
\begin{widetext}
\begin{equation}
    \frac{E(a,\alpha)}{A}=\frac{1}{4a}\dv[2]{\alpha} \sum\limits_{m=0}^\infty \frac{B_m}{m!}\left(\frac{\alpha\pi}{a}\right)^{m-2} + \frac{\kappa\eta^2}{4a}\dv[4]{\alpha} \sum\limits_{m=0}^\infty \frac{B_m}{m!}\left(\frac{\alpha\pi}{a}\right)^{m-2} \text{.}
\end{equation}
\end{widetext}
When we expand the series, we are left with some divergent quantities when we set $\alpha=0$ that we discard for $m<2$.
We can also see that the only relevant terms after differentiating and setting $\alpha=0$ are $m=4$ for the first terms, and $m=6$ for the second.
These depend upon $B_4=-\frac{1}{30}$ and $B_6=\frac{1}{42}$. 
Performing the differentiation, substituting in these values, and dividing by $a$, we arrive at our final result,
\begin{equation}\label{eqn:casimir_app_momres}
    \expval{T_{tt}} = \frac{-\pi^2}{1440 a^4}\left ( 1- \frac{12 \kappa \eta^2 \pi^2}{42 a^2}\right ) \text{.}
\end{equation}

Comparing this result to \Cref{eqn:casimir_app_momres}, we notice that the correction term is slightly larger than the point-splitting calculation, a result we address in \Cref{sec:casimir_density}.

\bibliography{PixelatedCasimir}

\begin{thebibliography}{27}%
\makeatletter
\providecommand \@ifxundefined [1]{%
 \@ifx{#1\undefined}
}%
\providecommand \@ifnum [1]{%
 \ifnum #1\expandafter \@firstoftwo
 \else \expandafter \@secondoftwo
 \fi
}%
\providecommand \@ifx [1]{%
 \ifx #1\expandafter \@firstoftwo
 \else \expandafter \@secondoftwo
 \fi
}%
\providecommand \natexlab [1]{#1}%
\providecommand \enquote  [1]{``#1''}%
\providecommand \bibnamefont  [1]{#1}%
\providecommand \bibfnamefont [1]{#1}%
\providecommand \citenamefont [1]{#1}%
\providecommand \href@noop [0]{\@secondoftwo}%
\providecommand \href [0]{\begingroup \@sanitize@url \@href}%
\providecommand \@href[1]{\@@startlink{#1}\@@href}%
\providecommand \@@href[1]{\endgroup#1\@@endlink}%
\providecommand \@sanitize@url [0]{\catcode `\\12\catcode `\$12\catcode
  `\&12\catcode `\#12\catcode `\^12\catcode `\_12\catcode `\%12\relax}%
\providecommand \@@startlink[1]{}%
\providecommand \@@endlink[0]{}%
\providecommand \url  [0]{\begingroup\@sanitize@url \@url }%
\providecommand \@url [1]{\endgroup\@href {#1}{\urlprefix }}%
\providecommand \urlprefix  [0]{URL }%
\providecommand \Eprint [0]{\href }%
\providecommand \doibase [0]{https://doi.org/}%
\providecommand \selectlanguage [0]{\@gobble}%
\providecommand \bibinfo  [0]{\@secondoftwo}%
\providecommand \bibfield  [0]{\@secondoftwo}%
\providecommand \translation [1]{[#1]}%
\providecommand \BibitemOpen [0]{}%
\providecommand \bibitemStop [0]{}%
\providecommand \bibitemNoStop [0]{.\EOS\space}%
\providecommand \EOS [0]{\spacefactor3000\relax}%
\providecommand \BibitemShut  [1]{\csname bibitem#1\endcsname}%
\let\auto@bib@innerbib\@empty
\bibitem [{\citenamefont {Hossenfelder}(2013)}]{hossenfelder2013minimal}%
  \BibitemOpen
  \bibfield  {author} {\bibinfo {author} {\bibfnamefont {S.}~\bibnamefont
  {Hossenfelder}},\ }\bibfield  {title} {\bibinfo {title} {Minimal length scale
  scenarios for quantum gravity},\ }\href@noop {} {\bibfield  {journal}
  {\bibinfo  {journal} {Living Reviews in Relativity}\ }\textbf {\bibinfo
  {volume} {16}},\ \bibinfo {pages} {1} (\bibinfo {year} {2013})}\BibitemShut
  {NoStop}%
\bibitem [{\citenamefont {Wheeler}(2018)}]{wheeler2018information}%
  \BibitemOpen
  \bibfield  {author} {\bibinfo {author} {\bibfnamefont {J.~A.}\ \bibnamefont
  {Wheeler}},\ }\href@noop {} {\bibinfo {title} {Information, physics, quantum:
  The search for links}} (\bibinfo {year} {2018})\BibitemShut {NoStop}%
\bibitem [{\citenamefont {Rovelli}\ and\ \citenamefont
  {Vidotto}(2014)}]{rovelli2014covariant}%
  \BibitemOpen
  \bibfield  {author} {\bibinfo {author} {\bibfnamefont {C.}~\bibnamefont
  {Rovelli}}\ and\ \bibinfo {author} {\bibfnamefont {F.}~\bibnamefont
  {Vidotto}},\ }\href@noop {} {\emph {\bibinfo {title} {Covariant loop quantum
  gravity: an elementary introduction to quantum gravity and spinfoam
  theory}}}\ (\bibinfo  {publisher} {Cambridge University Press},\ \bibinfo
  {year} {2014})\BibitemShut {NoStop}%
\bibitem [{\citenamefont {Trugenberger}(2015)}]{trugenberger2015quantum}%
  \BibitemOpen
  \bibfield  {author} {\bibinfo {author} {\bibfnamefont {C.~A.}\ \bibnamefont
  {Trugenberger}},\ }\bibfield  {title} {\bibinfo {title} {Quantum gravity as
  an information network self-organization of a 4d universe},\ }\href@noop {}
  {\bibfield  {journal} {\bibinfo  {journal} {Physical Review D}\ }\textbf
  {\bibinfo {volume} {92}},\ \bibinfo {pages} {084014} (\bibinfo {year}
  {2015})}\BibitemShut {NoStop}%
\bibitem [{\citenamefont {Tee}(2020)}]{tee2020dynamics}%
  \BibitemOpen
  \bibfield  {author} {\bibinfo {author} {\bibfnamefont {P.}~\bibnamefont
  {Tee}},\ }\bibfield  {title} {\bibinfo {title} {Dynamics and the emergence of
  geometry in an information mesh},\ }\href@noop {} {\bibfield  {journal}
  {\bibinfo  {journal} {The European Physical Journal C}\ }\textbf {\bibinfo
  {volume} {80}},\ \bibinfo {pages} {1} (\bibinfo {year} {2020})}\BibitemShut
  {NoStop}%
\bibitem [{\citenamefont {Dowker}(2006)}]{dowker2006causal}%
  \BibitemOpen
  \bibfield  {author} {\bibinfo {author} {\bibfnamefont {F.}~\bibnamefont
  {Dowker}},\ }\bibfield  {title} {\bibinfo {title} {Causal sets as discrete
  spacetime},\ }\href@noop {} {\bibfield  {journal} {\bibinfo  {journal}
  {Contemporary Physics}\ }\textbf {\bibinfo {volume} {47}},\ \bibinfo {pages}
  {1} (\bibinfo {year} {2006})}\BibitemShut {NoStop}%
\bibitem [{\citenamefont {Ambj}\ \emph {et~al.}(1997)\citenamefont {Ambj},
  \citenamefont {Ambj{\o}rn}, \citenamefont {Durhuus}, \citenamefont {Jonsson},
  \citenamefont {Jonsson} \emph {et~al.}}]{ambj1997quantum}%
  \BibitemOpen
  \bibfield  {author} {\bibinfo {author} {\bibfnamefont {J.}~\bibnamefont
  {Ambj}}, \bibinfo {author} {\bibfnamefont {J.}~\bibnamefont {Ambj{\o}rn}},
  \bibinfo {author} {\bibfnamefont {B.}~\bibnamefont {Durhuus}}, \bibinfo
  {author} {\bibfnamefont {T.}~\bibnamefont {Jonsson}}, \bibinfo {author}
  {\bibfnamefont {O.}~\bibnamefont {Jonsson}}, \emph {et~al.},\ }\href@noop {}
  {\emph {\bibinfo {title} {Quantum geometry: a statistical field theory
  approach}}}\ (\bibinfo  {publisher} {Cambridge University Press},\ \bibinfo
  {year} {1997})\BibitemShut {NoStop}%
\bibitem [{\citenamefont {Amelino-Camelia}(2005)}]{amelino2005introduction}%
  \BibitemOpen
  \bibfield  {author} {\bibinfo {author} {\bibfnamefont {G.}~\bibnamefont
  {Amelino-Camelia}},\ }\bibfield  {title} {\bibinfo {title} {Introduction to
  quantum-gravity phenomenology},\ }in\ \href@noop {} {\emph {\bibinfo
  {booktitle} {Planck Scale Effects in Astrophysics and Cosmology}}}\ (\bibinfo
   {publisher} {Springer},\ \bibinfo {year} {2005})\ pp.\ \bibinfo {pages}
  {59--100}\BibitemShut {NoStop}%
\bibitem [{\citenamefont {Amelino-Camelia}(2001)}]{amelino2001testable}%
  \BibitemOpen
  \bibfield  {author} {\bibinfo {author} {\bibfnamefont {G.}~\bibnamefont
  {Amelino-Camelia}},\ }\bibfield  {title} {\bibinfo {title} {Testable scenario
  for relativity with minimum length},\ }\href@noop {} {\bibfield  {journal}
  {\bibinfo  {journal} {Physics Letters B}\ }\textbf {\bibinfo {volume}
  {510}},\ \bibinfo {pages} {255} (\bibinfo {year} {2001})}\BibitemShut
  {NoStop}%
\bibitem [{\citenamefont {Amelino-Camelia}(2002)}]{amelino2002relativity}%
  \BibitemOpen
  \bibfield  {author} {\bibinfo {author} {\bibfnamefont {G.}~\bibnamefont
  {Amelino-Camelia}},\ }\bibfield  {title} {\bibinfo {title} {Relativity in
  spacetimes with short-distance structure governed by an observer-independent
  (planckian) length scale},\ }\href@noop {} {\bibfield  {journal} {\bibinfo
  {journal} {International Journal of Modern Physics D}\ }\textbf {\bibinfo
  {volume} {11}},\ \bibinfo {pages} {35} (\bibinfo {year} {2002})}\BibitemShut
  {NoStop}%
\bibitem [{\citenamefont {Amelino-Camelia}\ and\ \citenamefont
  {Arzano}(2002)}]{amelino2002coproduct}%
  \BibitemOpen
  \bibfield  {author} {\bibinfo {author} {\bibfnamefont {G.}~\bibnamefont
  {Amelino-Camelia}}\ and\ \bibinfo {author} {\bibfnamefont {M.}~\bibnamefont
  {Arzano}},\ }\bibfield  {title} {\bibinfo {title} {Coproduct and star product
  in field theories on lie-algebra noncommutative space-times},\ }\href@noop {}
  {\bibfield  {journal} {\bibinfo  {journal} {Physical Review D}\ }\textbf
  {\bibinfo {volume} {65}},\ \bibinfo {pages} {084044} (\bibinfo {year}
  {2002})}\BibitemShut {NoStop}%
\bibitem [{\citenamefont {Ho{\v{r}}ava}(2009)}]{hovrava2009quantum}%
  \BibitemOpen
  \bibfield  {author} {\bibinfo {author} {\bibfnamefont {P.}~\bibnamefont
  {Ho{\v{r}}ava}},\ }\bibfield  {title} {\bibinfo {title} {Quantum gravity at a
  lifshitz point},\ }\href@noop {} {\bibfield  {journal} {\bibinfo  {journal}
  {Physical Review D}\ }\textbf {\bibinfo {volume} {79}},\ \bibinfo {pages}
  {084008} (\bibinfo {year} {2009})}\BibitemShut {NoStop}%
\bibitem [{\citenamefont {Davies}\ and\ \citenamefont
  {Tee}(2023)}]{davies2023accelerated}%
  \BibitemOpen
  \bibfield  {author} {\bibinfo {author} {\bibfnamefont {P.~C.~W.}\
  \bibnamefont {Davies}}\ and\ \bibinfo {author} {\bibfnamefont
  {P.}~\bibnamefont {Tee}},\ }\bibfield  {title} {\bibinfo {title} {Accelerated
  particle detectors with modified dispersion relations},\ }\href@noop {}
  {\bibfield  {journal} {\bibinfo  {journal} {Physical Review D}\ }\textbf
  {\bibinfo {volume} {108}},\ \bibinfo {pages} {045009} (\bibinfo {year}
  {2023})}\BibitemShut {NoStop}%
\bibitem [{\citenamefont {Tee}\ and\ \citenamefont
  {Jafari}(2022)}]{tee2022fundamental}%
  \BibitemOpen
  \bibfield  {author} {\bibinfo {author} {\bibfnamefont {P.}~\bibnamefont
  {Tee}}\ and\ \bibinfo {author} {\bibfnamefont {N.}~\bibnamefont {Jafari}},\
  }\bibfield  {title} {\bibinfo {title} {Fundamental length scale and the
  bending of light in a gravitational field},\ }\href@noop {} {\bibfield
  {journal} {\bibinfo  {journal} {The European Physical Journal C}\ }\textbf
  {\bibinfo {volume} {82}},\ \bibinfo {pages} {1} (\bibinfo {year}
  {2022})}\BibitemShut {NoStop}%
\bibitem [{\citenamefont {Snyder}(1947)}]{snyder1947quantized}%
  \BibitemOpen
  \bibfield  {author} {\bibinfo {author} {\bibfnamefont {H.~S.}\ \bibnamefont
  {Snyder}},\ }\bibfield  {title} {\bibinfo {title} {Quantized space-time},\
  }\href@noop {} {\bibfield  {journal} {\bibinfo  {journal} {Physical Review}\
  }\textbf {\bibinfo {volume} {71}},\ \bibinfo {pages} {38} (\bibinfo {year}
  {1947})}\BibitemShut {NoStop}%
\bibitem [{\citenamefont {Amelino-Camelia}\ \emph {et~al.}(2016)\citenamefont
  {Amelino-Camelia}, \citenamefont {Gubitosi},\ and\ \citenamefont
  {Palmisano}}]{amelino2016pathways}%
  \BibitemOpen
  \bibfield  {author} {\bibinfo {author} {\bibfnamefont {G.}~\bibnamefont
  {Amelino-Camelia}}, \bibinfo {author} {\bibfnamefont {G.}~\bibnamefont
  {Gubitosi}},\ and\ \bibinfo {author} {\bibfnamefont {G.}~\bibnamefont
  {Palmisano}},\ }\bibfield  {title} {\bibinfo {title} {Pathways to
  relativistic curved momentum spaces: de sitter case study},\ }\href@noop {}
  {\bibfield  {journal} {\bibinfo  {journal} {International Journal of Modern
  Physics D}\ }\textbf {\bibinfo {volume} {25}},\ \bibinfo {pages} {1650027}
  (\bibinfo {year} {2016})}\BibitemShut {NoStop}%
\bibitem [{\citenamefont {Amelino-Camelia}\ \emph {et~al.}(2012)\citenamefont
  {Amelino-Camelia}, \citenamefont {Arzano}, \citenamefont {Kowalski-Glikman},
  \citenamefont {Rosati},\ and\ \citenamefont
  {Trevisan}}]{amelino2012relative}%
  \BibitemOpen
  \bibfield  {author} {\bibinfo {author} {\bibfnamefont {G.}~\bibnamefont
  {Amelino-Camelia}}, \bibinfo {author} {\bibfnamefont {M.}~\bibnamefont
  {Arzano}}, \bibinfo {author} {\bibfnamefont {J.}~\bibnamefont
  {Kowalski-Glikman}}, \bibinfo {author} {\bibfnamefont {G.}~\bibnamefont
  {Rosati}},\ and\ \bibinfo {author} {\bibfnamefont {G.}~\bibnamefont
  {Trevisan}},\ }\bibfield  {title} {\bibinfo {title} {Relative-locality
  distant observers and the phenomenology of momentum-space geometry},\
  }\href@noop {} {\bibfield  {journal} {\bibinfo  {journal} {Classical and
  Quantum Gravity}\ }\textbf {\bibinfo {volume} {29}},\ \bibinfo {pages}
  {075007} (\bibinfo {year} {2012})}\BibitemShut {NoStop}%
\bibitem [{\citenamefont {Casimir}\ and\ \citenamefont
  {Polder}(1948)}]{casimir1948influence}%
  \BibitemOpen
  \bibfield  {author} {\bibinfo {author} {\bibfnamefont {H.~B.}\ \bibnamefont
  {Casimir}}\ and\ \bibinfo {author} {\bibfnamefont {D.}~\bibnamefont
  {Polder}},\ }\bibfield  {title} {\bibinfo {title} {The influence of
  retardation on the london-van der waals forces},\ }\href@noop {} {\bibfield
  {journal} {\bibinfo  {journal} {Physical Review}\ }\textbf {\bibinfo {volume}
  {73}},\ \bibinfo {pages} {360} (\bibinfo {year} {1948})}\BibitemShut
  {NoStop}%
\bibitem [{\citenamefont {Brown}\ and\ \citenamefont
  {Maclay}(1969)}]{brown1969vacuum}%
  \BibitemOpen
  \bibfield  {author} {\bibinfo {author} {\bibfnamefont {L.~S.}\ \bibnamefont
  {Brown}}\ and\ \bibinfo {author} {\bibfnamefont {G.~J.}\ \bibnamefont
  {Maclay}},\ }\bibfield  {title} {\bibinfo {title} {Vacuum stress between
  conducting plates: an image solution},\ }\href@noop {} {\bibfield  {journal}
  {\bibinfo  {journal} {Physical Review}\ }\textbf {\bibinfo {volume} {184}},\
  \bibinfo {pages} {1272} (\bibinfo {year} {1969})}\BibitemShut {NoStop}%
\bibitem [{\citenamefont {Hong-Hao}\ \emph {et~al.}(2010)\citenamefont
  {Hong-Hao}, \citenamefont {Kai-Xi}, \citenamefont {Si-Wei}, \citenamefont
  {An},\ and\ \citenamefont {Xue-Song}}]{hong2010analytic}%
  \BibitemOpen
  \bibfield  {author} {\bibinfo {author} {\bibfnamefont {Z.}~\bibnamefont
  {Hong-Hao}}, \bibinfo {author} {\bibfnamefont {F.}~\bibnamefont {Kai-Xi}},
  \bibinfo {author} {\bibfnamefont {Q.}~\bibnamefont {Si-Wei}}, \bibinfo
  {author} {\bibfnamefont {Z.}~\bibnamefont {An}},\ and\ \bibinfo {author}
  {\bibfnamefont {L.}~\bibnamefont {Xue-Song}},\ }\bibfield  {title} {\bibinfo
  {title} {On analytic formulas of feynman propagators in position space},\
  }\href@noop {} {\bibfield  {journal} {\bibinfo  {journal} {Chinese Physics
  C}\ }\textbf {\bibinfo {volume} {34}},\ \bibinfo {pages} {1576} (\bibinfo
  {year} {2010})}\BibitemShut {NoStop}%
\bibitem [{\citenamefont {Birrell}\ and\ \citenamefont
  {Davies}(1984)}]{birrell1984quantum}%
  \BibitemOpen
  \bibfield  {author} {\bibinfo {author} {\bibfnamefont {N.~D.}\ \bibnamefont
  {Birrell}}\ and\ \bibinfo {author} {\bibfnamefont {P.}~\bibnamefont
  {Davies}},\ }\href@noop {} {\emph {\bibinfo {title} {Quantum fields in curved
  space}}}\ (\bibinfo  {publisher} {Cambridge university press},\ \bibinfo
  {year} {1984})\BibitemShut {NoStop}%
\bibitem [{\citenamefont {Gradshteyn}\ and\ \citenamefont
  {Ryzhik}(2014)}]{gradshteyn2014table}%
  \BibitemOpen
  \bibfield  {author} {\bibinfo {author} {\bibfnamefont {I.~S.}\ \bibnamefont
  {Gradshteyn}}\ and\ \bibinfo {author} {\bibfnamefont {I.~M.}\ \bibnamefont
  {Ryzhik}},\ }\href@noop {} {\emph {\bibinfo {title} {Table of integrals,
  series, and products}}}\ (\bibinfo  {publisher} {Academic press},\ \bibinfo
  {year} {2014})\BibitemShut {NoStop}%
\bibitem [{\citenamefont {Fierz}(1959)}]{fierz1959anziehung}%
  \BibitemOpen
  \bibfield  {author} {\bibinfo {author} {\bibfnamefont {M.}~\bibnamefont
  {Fierz}},\ }\href@noop {} {\emph {\bibinfo {title} {Zur anziehung leitender
  ebenen im vakuum}}},\ \bibinfo {type} {Tech. Rep.}\ (\bibinfo  {institution}
  {CM-P00057201},\ \bibinfo {year} {1959})\BibitemShut {NoStop}%
\bibitem [{\citenamefont {Peskin}\ and\ \citenamefont
  {Schroeder}(1995)}]{peskin1995introduction}%
  \BibitemOpen
  \bibfield  {author} {\bibinfo {author} {\bibfnamefont {M.~E.}\ \bibnamefont
  {Peskin}}\ and\ \bibinfo {author} {\bibfnamefont {D.~V.}\ \bibnamefont
  {Schroeder}},\ }\href@noop {} {\emph {\bibinfo {title} {An introduction to
  quantum field theory}}}\ (\bibinfo  {publisher} {Perseus Books, Reading MA},\
  \bibinfo {year} {1995})\BibitemShut {NoStop}%
\bibitem [{\citenamefont {Ford}(1976)}]{ford1976quantum}%
  \BibitemOpen
  \bibfield  {author} {\bibinfo {author} {\bibfnamefont {L.}~\bibnamefont
  {Ford}},\ }\bibfield  {title} {\bibinfo {title} {Quantum vacuum energy in a
  closed universe},\ }\href@noop {} {\bibfield  {journal} {\bibinfo  {journal}
  {Physical Review D}\ }\textbf {\bibinfo {volume} {14}},\ \bibinfo {pages}
  {3304} (\bibinfo {year} {1976})}\BibitemShut {NoStop}%
\bibitem [{\citenamefont {Rosati}(2024)}]{rosati2024private}%
  \BibitemOpen
  \bibfield  {author} {\bibinfo {author} {\bibfnamefont {G.}~\bibnamefont
  {Rosati}},\ }\href@noop {} {}\bibinfo {howpublished} {personal communication}
  (\bibinfo {year} {2024})\BibitemShut {NoStop}%
\bibitem [{\citenamefont {Amelino-Camelia}\ \emph {et~al.}(2011)\citenamefont
  {Amelino-Camelia}, \citenamefont {Freidel}, \citenamefont
  {Kowalski-Glikman},\ and\ \citenamefont {Smolin}}]{amelino2011principle}%
  \BibitemOpen
  \bibfield  {author} {\bibinfo {author} {\bibfnamefont {G.}~\bibnamefont
  {Amelino-Camelia}}, \bibinfo {author} {\bibfnamefont {L.}~\bibnamefont
  {Freidel}}, \bibinfo {author} {\bibfnamefont {J.}~\bibnamefont
  {Kowalski-Glikman}},\ and\ \bibinfo {author} {\bibfnamefont {L.}~\bibnamefont
  {Smolin}},\ }\bibfield  {title} {\bibinfo {title} {Principle of relative
  locality},\ }\href@noop {} {\bibfield  {journal} {\bibinfo  {journal}
  {Physical Review D}\ }\textbf {\bibinfo {volume} {84}},\ \bibinfo {pages}
  {084010} (\bibinfo {year} {2011})}\BibitemShut {NoStop}%
\end{thebibliography}%

\end{document}